\documentclass[conference]{IEEEtran}
\IEEEoverridecommandlockouts

\usepackage{cite}
\usepackage{amsmath,amssymb,amsfonts}
\usepackage{algorithmic}
\usepackage{algorithm}
\usepackage{graphicx}
\usepackage{textcomp}
\usepackage{xcolor}
\usepackage{subfigure}

\def\BibTeX{{\rm B\kern-.05em{\sc i\kern-.025em b}\kern-.08em
    T\kern-.1667em\lower.7ex\hbox{E}\kern-.125emX}}
\begin{document}

\newcommand{\toolname}[0]{}
\renewcommand{\toolname}[0]{MOLHEC }
\def\BibTeX{{\rm B\kern-.05em{\sc i\kern-.025em b}\kern-.08em
    T\kern-.1667em\lower.7ex\hbox{E}\kern-.125emX}}

\title{Multi-Objective Load Balancing for Heterogeneous Edge-Based Object Detection Systems
\thanks{This research was partially supported by the Australian Research Council (ARC) through funded projects DP230100081 and LP210200213.}
}

\author{
\IEEEauthorblockN{
Daghash K. Alqahtani$^{1}$,
Maria A. Rodriguez$^{2}$,
Muhammad Aamir Cheema$^{3}$,
Adel N. Toosi$^{1}$
}
\IEEEauthorblockA{
$^{1}$Distributed Systems and Network Applications (DisNet) Laboratory, The University of Melbourne, Australia\\
$^{2}$School of Computing and Information Systems, The University of Melbourne, Australia\\
$^{3}$Department of Software Systems and Cybersecurity, Monash University, Australia\\
\{daghash.alqahtani@student, maria.read, adel.toosi\}@unimelb.edu.au,
\{aamir.cheema\}@monash.edu
}
}

\maketitle

\begin{abstract}
The rapid proliferation of the Internet of Things (IoT) and smart applications has led to a surge in data generated by distributed sensing devices. Edge computing is a mainstream approach to managing this data by pushing computation closer to the data source, typically onto resource-constrained devices such as single-board computers (SBCs). In such environments, the unavoidable heterogeneity of hardware and software makes effective load balancing particularly challenging. In this paper, we propose a multi-objective load balancing method tailored to heterogeneous, edge-based object detection systems. We study a setting in which multiple device–model pairs expose distinct accuracy, latency, and energy profiles, while both request intensity and scene complexity fluctuate over time. To handle this dynamically varying environment, our approach uses a two-stage decision mechanism: it first performs accuracy-aware filtering to identify suitable device–model candidates that provide accuracy within the acceptable range, and then applies a weighted-sum scoring function over expected latency and energy consumption to select the final execution target. We evaluate the proposed load balancer through extensive experiments on real-world datasets, comparing against widely used baseline strategies. The results indicate that the proposed multi-objective load balancing method halves energy consumption and achieves an 80\% reduction in end-to-end latency, while incurring only a modest, up to 10\%, decrease in detection accuracy relative to an accuracy-centric baseline.

\end{abstract}

\begin{IEEEkeywords}
Edge computing, Load balancing, Object detection, Energy Consumption, Latency, Accuracy
\end{IEEEkeywords}

\section{Introduction}

The rapid expansion of the Internet of Things (IoT) and smart applications is driving a substantial increase in the scale and heterogeneity of data generated by distributed sensing devices. Recent industry forecasts estimate that connected IoT devices will grow from around 21 billion in 2025 to more than 38 billion by 2030~\cite{iotanalytics2025devices}. Much of this data growth is driven by camera-enabled endpoints that continuously produce high-rate image and video streams. Processing such data in centralized cloud infrastructures is often impractical due to bandwidth constraints and increased end-to-end latency, challenges that are particularly acute for video analytics where raw streams are data-intensive and contain sensitive visual content. Consequently, edge computing has become a norm for performing inference and analytics close to the data source on resource-constrained platforms such as Raspberry Pi, Jetson Orin Nano, and similar edge systems. This reduces communication overhead while improving responsiveness and scalability.

Object detection in vision-enabled IoT systems is a core component of many smart applications and has been widely adopted in domains such as agriculture, sports analytics, transportation, healthcare, and smart cities. Recent advances in machine learning and artificial intelligence have led to powerful object detection models that achieve high accuracy and can operate at or near real time. Representative families include You Only Look Once (YOLO)~\cite{Redmon2016}, Single Shot Multibox Detector (SSD)~\cite{Liu2016}, and EfficientDet~\cite{Tan2020}. These models are typically implemented using deep neural network architectures and are offered in multiple variants to support different computational budgets. For example, the YOLOv8 family~\cite{ultralytics} includes \textit{nano}, \textit{small}, \textit{medium}, and \textit{large} versions, each reflecting different trade-offs between model size, speed, and accuracy. Deploying such models on edge platforms has become a practical strategy to reduce latency, preserve privacy, and enable scalable on-device intelligence.

In edge computing environments, heterogeneity is essentially unavoidable. As systems evolve over time, new devices are added, existing hardware is upgraded or replaced, and new machine learning models are introduced. Under these conditions, maintaining a fully homogeneous platform in terms of hardware and software is difficult, if not impossible. This heterogeneity introduces fundamental challenges for resource management, since devices differ in computational capability, energy consumption, and model performance. Simple strategies that treat all nodes as identical are no longer adequate when some devices are more powerful but energy hungry, while others are more constrained but considerably more energy efficient.

This tension is particularly acute in scenarios where workloads and scene complexity (the number of objects in the video stream) vary over time and where multiple edge devices operate under strict energy budgets (e.g., smart city surveillance or precision agriculture), especially when they are powered by batteries or renewable sources like solar panels. In these settings, it is essential to prevent overloading individual edge nodes, limit latency, and conserve energy, while still providing reliable detection quality for safety critical applications such as surveillance, traffic monitoring, or industrial automation. These requirements motivate the need for load balancing mechanisms that can exploit device and model heterogeneity in a principled way and that can adapt to dynamic workload conditions.

In this work, we address this challenge by focusing on load balancing for heterogeneous edge-based object detection systems. Rather than assuming a homogeneous cluster or static task allocation decisions, we consider a setting in which multiple device and model pairs exhibit different accuracy, latency, and energy profiles, and where the workload and scene complexity vary over time. Our goal is to design a load balancer that can allocate tasks on a per-request basis, jointly considering three key objectives: energy consumption, latency, and accuracy. To this end, we propose a multi-objective load balancer for heterogeneous edge computing. The load balancer uses offline profiling to characterise each device and model pair, and then applies an accuracy-aware filtering step followed by multi-objective scoring at run time. The scoring stage incorporates both expected latency and energy consumption and uses a weighted-sum formulation to resolve trade-offs between these objectives. By integrating model accuracy, device heterogeneity, and dynamic workload information into a unified load-balancing strategy, our load balancer aims to improve the efficiency and robustness of smart city edge deployments.

The main contributions of this work are summarised as follows:
\begin{enumerate}
    \item We introduce a multi-objective load balancer for heterogeneous edge computing, together with a holistic system architecture that integrates vision enabled IoT devices, a gateway, and diverse edge nodes hosting different device and model combinations for smart city workloads.

    \item We formulate the multi objective load balancing problem for heterogeneous edge based object detection and design a two-stage decision mechanism that combines accuracy aware filtering with a weighted sum scoring function over expected latency and energy consumption.

    \item We implement the proposed load balancer on a heterogeneous edge testbed and present an experimental evaluation using a practical surveillance video stream under multiple workload conditions, comparing it against several baseline strategies. At high concurrency, the proposed configuration reduces average response time by more than 80 percent and halves the energy consumption per request relative to an accuracy oriented baseline, while maintaining accuracy within roughly 10 percent of that baseline.
\end{enumerate}


\section{Motivation}
\label{sec:motivation}
\begin{figure}[t]

    \subfigure[]{
        \includegraphics[width=0.46\columnwidth]{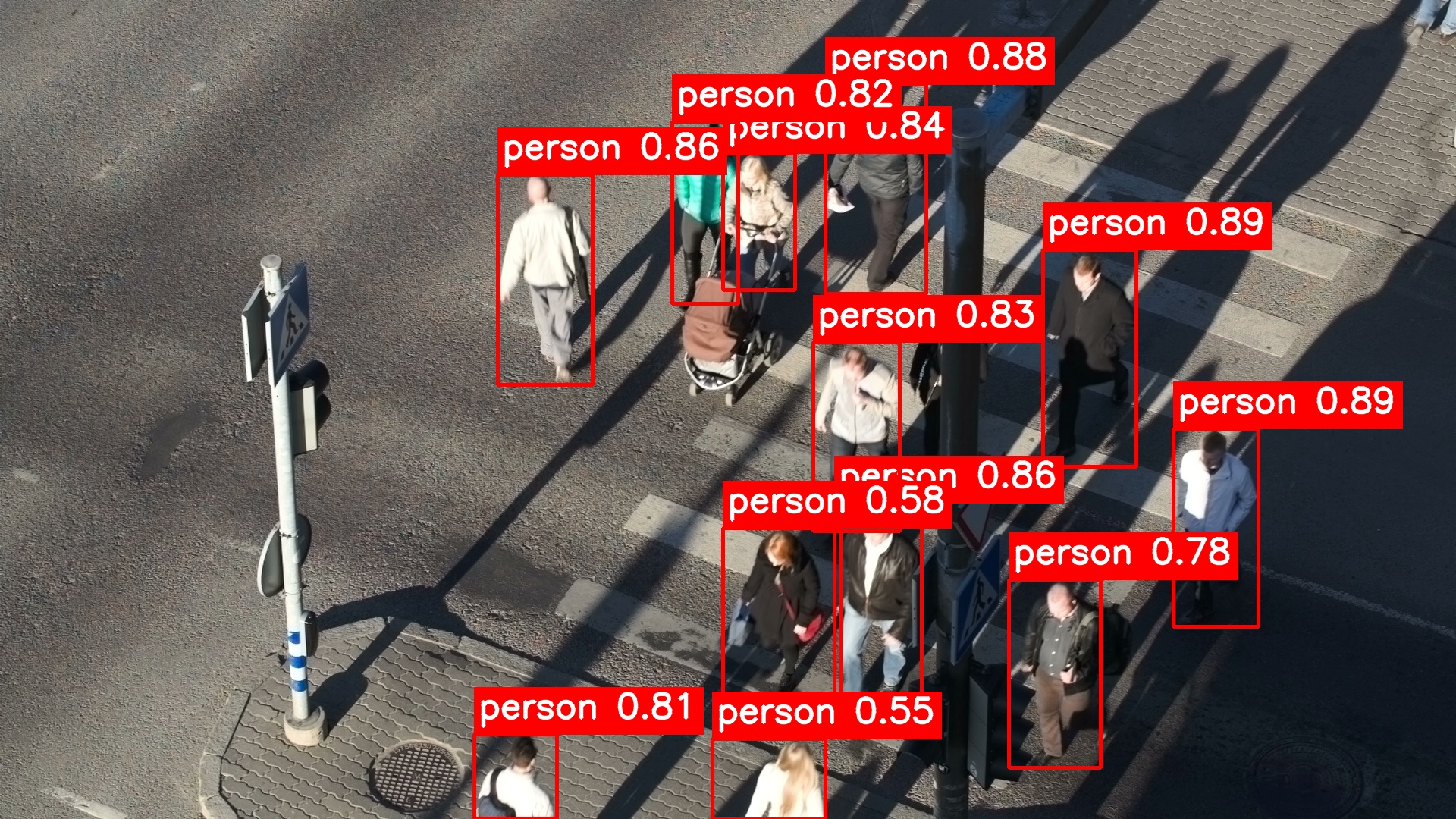} 
    }
  \subfigure[]{
        \includegraphics[width=0.46\columnwidth]{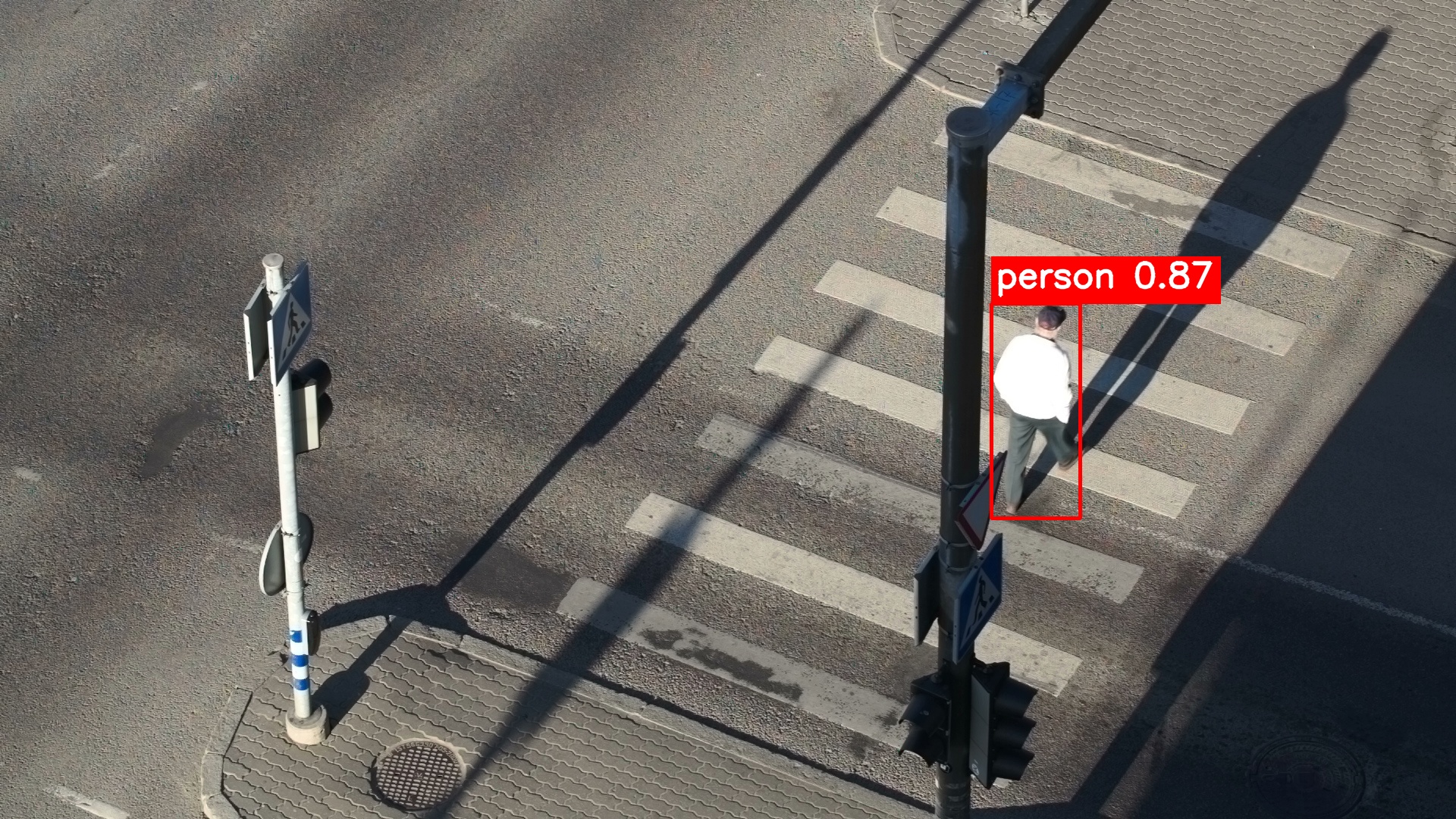}
    }
\vspace{-0.2cm}
\caption{Contrasting pedestrians scenarios: (a) High density scene; (b) Low density scene.}
\label{fig:scene_examples}
\end{figure}

Consider a road intersection where the density of pedestrians varies significantly over time, ranging from heavy congestion during rush hours to sparse activity during off-peak periods, as shown in Fig.~\ref{fig:scene_examples}. In high-density scenarios, maintaining high detection accuracy is critical for safety-related applications; therefore, the system should preferentially select device-model combinations that provide the best accuracy, even at a higher energy cost. In contrast, during periods of low pedestrian density, energy can be conserved by processing frames on device–model pairs that consume less power, while still achieving comparable accuracy.

At the same time, when the workload is high, it is not practical to process all frames on a single device and model pair. Concentrating requests on one node can lead to overload, queue buildup, larger response times, and degradation of quality of service. Consequently, there is a clear need for load balancing strategies that explicitly account for heterogeneity in edge computing systems and jointly consider energy consumption, latency, and accuracy. Designing mechanisms that can adapt to changing workload conditions and application requirements, while exploiting the diversity of device and model capabilities, is essential to ensure efficient and reliable operation of smart city edge deployments.

\begin{figure}[t]
    \centering
    \includegraphics[width=0.95\linewidth]{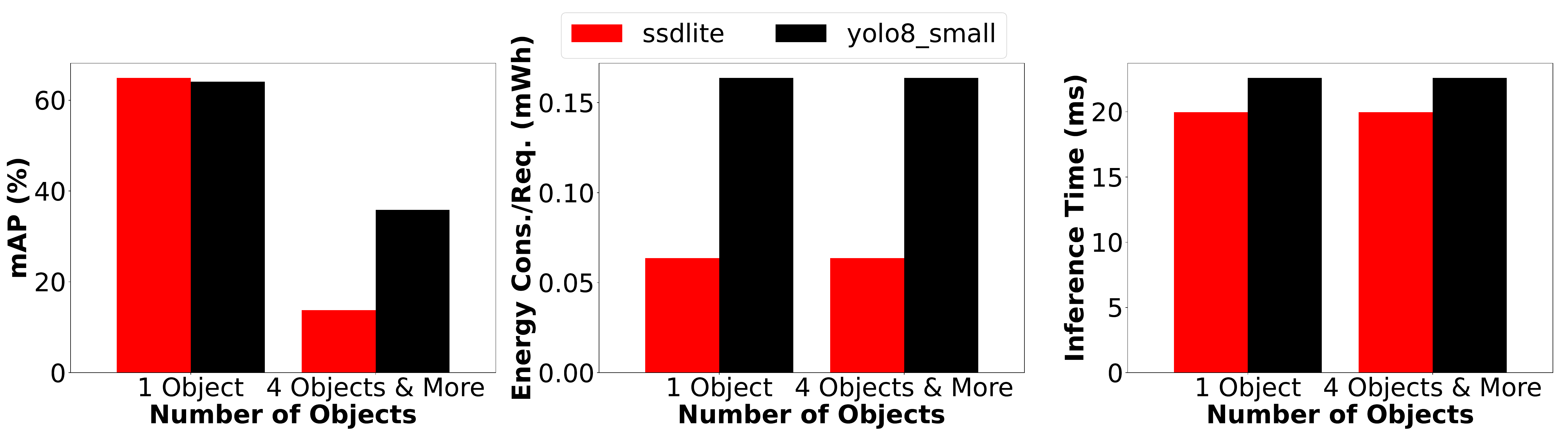}
    \vspace{-0.2cm}
    \caption{Accuracy, energy consumption and inference time for different group
of images across object detection models.}
    \vspace{-0.2cm}
    \label{fig:combined_group_results}
\end{figure}

To further illustrate these trade-offs, we conducted a preliminary experiment with two image groups: one containing a single object per image and another with four or more objects. For each group, we computed detection accuracy using mean Average Precision (mAP), a widely used object detection metric based on the area under the precision–recall curve, along with energy consumption and per image inference time. As shown in Fig.~\ref{fig:combined_group_results}, \textit{SSD Lite} and \textit{YOLOv8\_small} achieve very similar accuracy for single-object images, whereas \textit{YOLOv8\_small} nearly doubles the mAP of \textit{SSD Lite} on images with four or more objects. \textit{SSD Lite}, however, maintains a consistent energy cost across both groups that is approximately fifty percent lower than \textit{YOLOv8\_small} and reduces processing time by about 5 ms per request. These results indicate that in less complex scenes two object detection models can offer comparable accuracy, while substantial energy and latency gains can be obtained by selecting an efficient model such as \textit{SSD Lite}, whereas in more complex scenes a higher-capacity model like \textit{YOLOv8\_small} is justified by its clear accuracy advantage despite higher energy and slightly longer inference time. This preliminary evidence supports the need for task allocation strategies that adapt device-model selection to scene complexity, energy consumption, and inference time, rather than relying on a single static configuration.

\section{System Design and Implementation}
\label{sec:system_design}

\begin{figure}[t]
    \centering
    \includegraphics[width=0.95\linewidth]{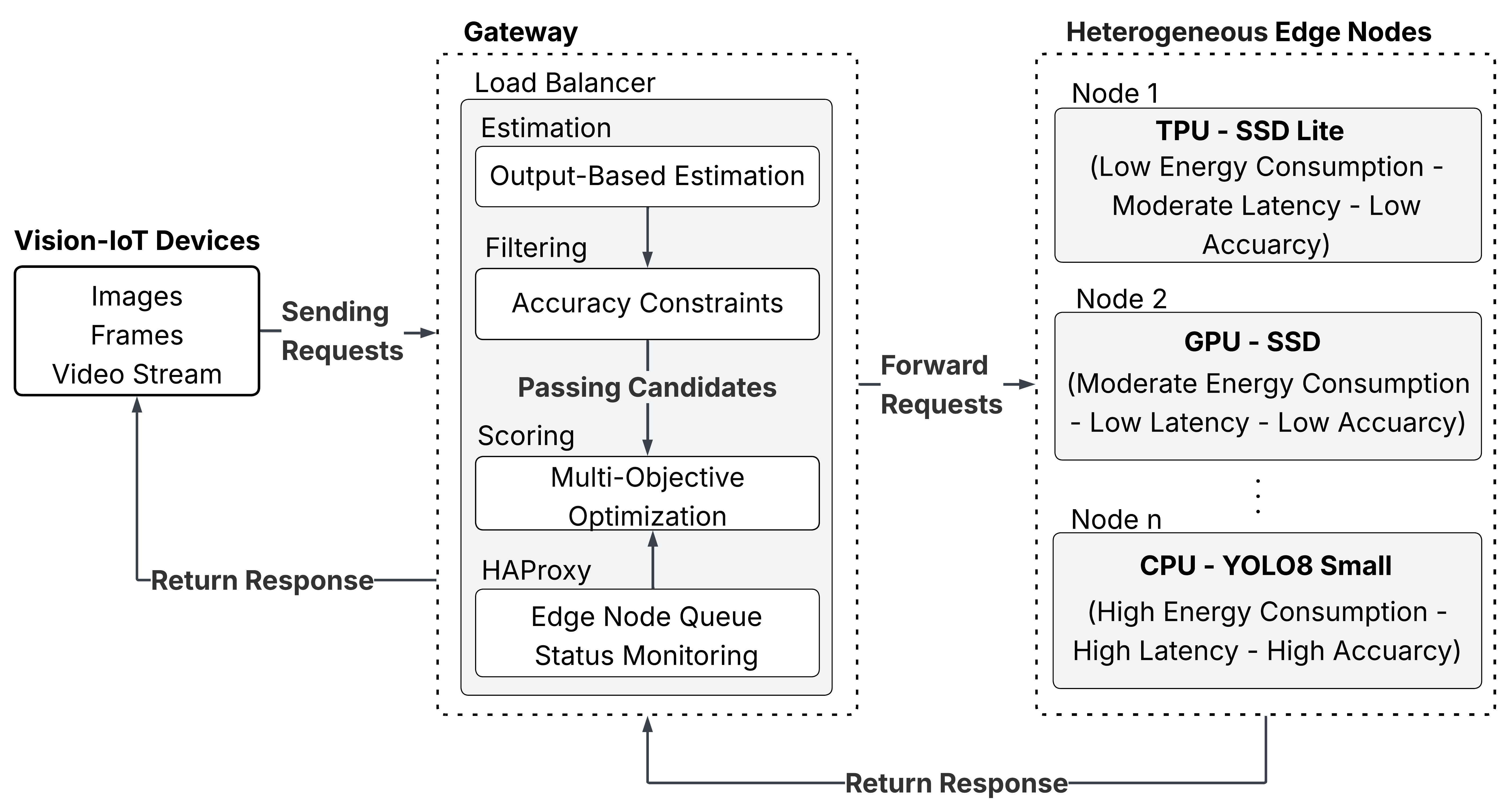}
    \vspace{-0.3cm}
    \caption{The overall system architecture comprises vision-enabled IoT devices, a gateway, and heterogeneous edge computing resources.}
    \vspace{-0.2cm}
    \label{fig:system-architecture}
\end{figure}

The system architecture, illustrated in Fig.~\ref{fig:system-architecture}, follows a layered design that reflects the end to end processing pipeline. At the IoT layer, vision-enabled devices capture images, frames, or video streams from the environment and send object detection requests to the gateway. The gateway layer hosts the proposed multi-objective load balancer, which implements the decision logic for task allocation. Inside the load balancer, an object count estimator first predicts the number of objects in each incoming image using an outputs based technique. The estimated object count is then passed to an accuracy based filtering component that applies the specified accuracy constraints and selects the subset of device-model pairs that satisfy the required tolerance.

The remaining candidates are forwarded to the scoring stage, where a multi-objective optimisation combines expected latency and energy consumption through a weighted sum. Expected latency is obtained using both offline profiling and real time queue information exposed by the HAProxy component, which monitors the status of all edge nodes. Based on the resulting scores, the load balancer selects an appropriate edge node and forwards the request to the heterogeneous edge layer. This layer consists of multiple edge nodes with different hardware architectures, for example central processing unit (CPU), graphics processing unit (GPU), and tensor processing unit (TPU), and different object detection models such as SSD, SSD Lite, and YOLOv8 small. The resulting device-model pairs exhibit diverse performance profiles, with some offering lower inference time at reduced accuracy, others providing better energy efficiency, and others achieving higher accuracy at higher energy and latency cost. Requests are processed on the selected edge node and the detection results are returned to the gateway, which completes the response to the vision-enabled IoT device.

\subsection{Problem Formulation}

Modern edge-based object detection systems at the edge frequently experience sudden spikes in incoming requests during peak hours, large public events, or emergencies. When the workload increases rapidly, it is essential to balance requests across heterogeneous edge nodes rather than overloading a single device. Overloading one node significantly increases end-to-end latency, which is unacceptable for latency-sensitive real-time applications such as pedestrian detection for autonomous vehicles, anomaly detection in public safety systems, emergency vehicle monitoring, etc. Concentrating requests on a single node also accelerates its energy depletion, especially when it is powered by batteries or renewable sources, which can cause it to drop out of service and reduce overall Quality of Service. At the same time, task allocation decisions must preserve accuracy, particularly for mission-critical applications such as detecting vulnerable road users, intrusion detection in restricted zones, and safety monitoring in industrial environments.

To formalize the load balancing problem, let $\mathcal{D}$ denote the set of edge devices, $\mathcal{M}$ the set of deployed object detection models, and $\mathcal{G}$ the set of image groups, where each group corresponds to a range of object counts in a frame. In our experiments, $\mathcal{G}$ comprises five groups, images with zero, one, two, three, and four or more objects.
 Let $\mathcal{P} \subseteq \mathcal{D} \times \mathcal{M}$ denote the set of device--model pairs. Offline profiling provides, for each pair $p \in \mathcal{P}$ and group $g \in \mathcal{G}$:
\begin{itemize}
    \item $T_{p,g}$: inference time per request (millisecond).
    \item $E_{p,g}$: energy consumption per request (mWh), excluding base device power.
    \item $\mathrm{mAP}_{p,g}$: accuracy for group $g$.
\end{itemize}

For a request belonging to a group $g$, the goal is to \textit{minimize expected latency} and \textit{minimize energy consumption}, subject to an application specific accuracy tolerance $\Delta_{\mathrm{mAP}}$ specified by the system designer or operator. Here $\Delta_{\mathrm{mAP}}$ represents the maximum allowable reduction in mAP relative to the best profiled device and model pair for group $g$. For example, a small value of $\Delta_{\mathrm{mAP}}$ permits only a few mAP points of loss and therefore favors high accuracy at the cost of higher energy and latency, whereas a larger value allows greater accuracy loss in exchange for selecting more energy efficient or faster configurations. In practice, $\Delta_{\mathrm{mAP}}$ is chosen according to application requirements, with safety critical deployments using tight tolerances and less critical analytics allowing larger tolerances to obtain greater energy savings.
Define the best available accuracy for group $g$ as
\[
\mathrm{mAP}^{\max}_g = \max_{p\in\mathcal{P}} \mathrm{mAP}_{p,g}.
\]

Given $\Delta_{\mathrm{mAP}}$, the minimum acceptable accuracy threshold for group $g$ is
\[
\mathrm{mAP}^{\mathrm{thr}}_g = \mathrm{mAP}^{\max}_g - \Delta_{\mathrm{mAP}}.
\]

The accuracy-feasible set of candidate pairs is then
\[
\mathcal{C}_g(\Delta_{\mathrm{mAP}})
=
\left\{
    p \in \mathcal{P} : 
    \mathrm{mAP}_{p,g} \ge \mathrm{mAP}^{\mathrm{thr}}_g
\right\}.
\]

At run time, the gateway queries the underlying load balancing component to obtain the instantaneous request queue length $q_p$ for each device-model pair $p$. The expected latency is approximated as
\[
L_{p,g}^{\mathrm{exp}} \approx T_{p,g} \times\,(1+q_p),
\]
where $T_{p,g}$ is the profiled inference time for group $g$ on candidate $p$, and $q_p$ is the number of pending requests ahead of the arriving request at $p$ (with $q_p=0$ indicating an idle server).

The resulting multi-objective optimization problem is
\[
\min_{p \in \mathcal{C}_g(\Delta_{\mathrm{mAP}})}
\left(
    L_{p,g}^{\mathrm{exp}},\;
    E_{p,g}
\right),
\]
i.e., selecting a device--model pair that simultaneously reduces expected
latency and energy consumption over the accuracy-feasible set
$\mathcal{C}_g(\Delta_{\mathrm{mAP}})$. This formulation makes explicit the
inherent latency--energy trade-off in heterogeneous edge environments under
dynamic workloads while maintaining the required accuracy constraint.
As detailed in the next section, we implement this multi-objective criterion through a weighted sum scalarization over latency and energy, enabling an online selection rule that is computationally efficient at inference time. Rather than scalarising all three metrics jointly, we treat accuracy as a hard constraint and enforce an application specific tolerance on mAP, which guarantees a minimum detection quality and prevents the load balancer from trading excessive accuracy losses for modest gains in latency or energy.

\subsection{Proposed Solution}

To efficiently solve the multi-objective task allocation problem in real time, we propose a two-stage load-balancing mechanism consisting of: (i) accuracy-based filtering, and (ii) multi-objective scoring using a weighted-sum method. This mechanism is executed inside the API gateway via HAProxy~\footnote{https://www.haproxy.com/} and Lua script~\footnote{https://www.lua.org/about.html}, enabling microsecond-scale decision latency and efficient request dispatching. The overall decision process is summarized in Algorithm~\ref{alg:load_balancing}.

\subsubsection{Stage 1: Accuracy-Based Filtering}

This step ensures that task allocation decisions do not compromise accuracy, which is particularly important for safety-critical applications.
Given an incoming request, the object count estimator assigns it to an image group $g$ using an output based method that exploits temporal continuity in image streams. For video, the estimator assumes that consecutive frames contain a similar number of objects, so instead of running an additional counting model, the system reuses the object count obtained from the detection output of the previous frame produced by the last selected device-model pair. This output based strategy avoids per frame estimation, keeps the additional computation overhead negligible, and improves energy efficiency compared with using a separate counting model. A detailed description and evaluation of this estimator is provided in \cite{alqahtani2025}. The load balancer then considers only the accuracy-feasible candidates $\mathcal{C}_g(\Delta_{\mathrm{mAP}})$, as defined above, i.e., device--model pairs whose $\mathrm{mAP}_{p,g}$ meets the previously specified threshold. This filtering ensures that subsequent task-allocation decisions preserve detection accuracy, which is essential in safety-critical deployments.

\subsubsection{Stage 2: Multi-Objective Scoring via Weighted-Sum Method}

For each candidate $p \in \mathcal{C}_g(\Delta_{\mathrm{mAP}})$, the expected latency $L_{p,g}^{\mathrm{exp}}$ is computed using real-time queue information. To enable a fair comparison and aggregation of latency and energy despite their different units and scales, both metrics are normalized over the candidate set $\mathcal{C}_g(\Delta_{\mathrm{mAP}})$, where $L_g^{\min}=\min_{p} L_{p,g}^{\mathrm{exp}}$ and $L_g^{\max}=\max_{p} L_{p,g}^{\mathrm{exp}}$, and $E_g^{\min}=\min_{p} E_{p,g}$ and $E_g^{\max}=\max_{p} E_{p,g}$, with all extrema taken over $p \in \mathcal{C}_g(\Delta_{\mathrm{mAP}})$. The normalized metrics are computed as
\[
\tilde{L}_{p,g} =
\frac{L_{p,g}^{\mathrm{exp}} - L_g^{\min}}
     {L_g^{\max} - L_g^{\min}},
\qquad
\tilde{E}_{p,g} =
\frac{E_{p,g} - E_g^{\min}}
     {E_g^{\max} - E_g^{\min}}.
\]

To combine the competing objectives of latency and energy, we adopt the \textit{weighted-sum} approach, one of the most commonly used scalarization methods in multi-objective optimization. This method is widely preferred because it offers:
(i) low computational overhead,
(ii) intuitive control over objective trade-offs via a single weight, and
(iii) compatibility with online per-request decision making.

The scalarized score is
\[
J_{p,g} = 
\gamma \tilde{L}_{p,g}
+
(1 - \gamma)\tilde{E}_{p,g},
\]
where $\gamma \in [0,1]$ determines the emphasis on latency versus energy. The selected pair is
\[
p^\star(g)
=
\arg\min_{p\in\mathcal{C}_g(\Delta_{\mathrm{mAP}})}
J_{p,g}.
\]

\begin{algorithm}[t]
\caption{Multi-Objective Load Balancer Algorithm}
\begin{algorithmic}[1]

\REQUIRE
$g$ (image group), accuracy tolerance $\Delta_{\mathrm{mAP}}$, 
profiling data, 
queue states $q[p]$, 
weight $\gamma$

\ENSURE
$p^\star$ --- selected device--model pair

\STATE $\mathrm{mAP}_{\max} \gets \max_{p\in\mathcal{P}} \mathrm{mAP}[p,g]$
\STATE $\mathrm{mAP}_{\mathrm{thr}} \gets \mathrm{mAP}_{\max} - \Delta_{\mathrm{mAP}}$
\STATE $\mathcal{C} \gets 
\{\, p \in \mathcal{P} : 
\mathrm{mAP}[p,g] \ge \mathrm{mAP}_{\mathrm{thr}} \,\}$

\FOR{$p \in \mathcal{C}$}

    \STATE $L_{\exp}[p] \gets T[p,g] \cdot (1 + q[p])$
\ENDFOR

\STATE $L_{\min}, L_{\max} \gets \min/\max_{p \in \mathcal{C}} L_{\exp}[p]$
\STATE $E_{\min}, E_{\max} \gets \min/\max_{p \in \mathcal{C}} E[p,g]$

\FOR{$p \in \mathcal{C}$}
    \STATE $L_{\text{norm}}[p] \gets (L_{\exp}[p] - L_{\min}) / (L_{\max} - L_{\min})$
    \STATE $E_{\text{norm}}[p] \gets (E[p,g] - E_{\min}) / (E_{\max} - E_{\min})$
    \STATE $J[p] \gets \gamma L_{\text{norm}}[p] + (1-\gamma) E_{\text{norm}}[p]$
\ENDFOR

\RETURN $\arg\min_{p\in\mathcal{C}} J[p]$

\end{algorithmic}
\label{alg:load_balancing}
\end{algorithm}

\section{Performance evaluation}
\label{sec:performance_evaluation}

\subsection{Experimental Setup}
\subsubsection{Dataset}

For our experimental evaluation of the proposed load balancer and all baseline methods, we used a real-world video stream capturing pedestrians crossing a road,\footnote{https://www.vecteezy.com/video/28257902-people-crossing-the-street-on-green-light} which we treated as an object detection dataset. Since no manual annotations were available, we first decoded the raw video and extracted every frame at the original frame rate, obtaining a pool of 417 JPEG images. To construct reference labels, we then applied the high-capacity YOLOv8x model to each image and recorded its detections in YOLO format as a proxy for ground-truth annotations. These pseudo ground-truth annotations were subsequently used to assess the detection performance of different device and model pairs under the task assignment decisions produced by the proposed load balancer and the baseline load balancing strategies.

\subsubsection{Experimental Testbed}

This section describes the experimental testbed used to evaluate the proposed load balancing approach. The testbed builds on an extensive benchmarking campaign that systematically evaluated combinations of edge devices and object detection models with respect to three key metrics, energy consumption, inference time, and mean Average Precision (mAP)~\cite{daghash2024}. The benchmarking covered eight heterogeneous SBC edge devices, \textit{Raspberry Pi 3}, \textit{Raspberry Pi 4}, \textit{Raspberry Pi 5}, and the \textit{Jetson Orin Nano}, with the Raspberry Pi boards evaluated both \textit{with} and \textit{without TPU} and, for \textit{Raspberry Pi 5}, also with an \textit{AI HAT}. On these platforms, we evaluated eight object detection configurations, \textit{SSD v1}, \textit{SSD Lite}, \textit{EfficientDet Lite0}, \textit{EfficientDet Lite1}, \textit{EfficientDet Lite2}, \textit{YOLOv8 nano}, \textit{YOLOv8 small}, and \textit{YOLOv8 medium}. All device and model pairs were profiled across five object count groups, images with \textit{zero}, \textit{one}, \textit{two}, \textit{three}, and \textit{four or more objects}, yielding a dataset that captures how latency, energy, and accuracy vary across device and model pairs under different levels of scene complexity. From this dataset, we retain only configurations that offer meaningful trade offs, where each selected pair is strong in at least one metric, and discard consistently inferior combinations. The subset of device, model, runtime, and target metric combinations used in our evaluation is summarised in Table~\ref{tab:testbed_pairs}.

\begin{table}[t]
\scriptsize
\centering
\caption{Selected device and model configurations for edge nodes.}
\label{tab:testbed_pairs}
\begin{tabular}{|c|c|c|c|}
\hline
\textbf{Metric / Group} & \textbf{Edge Device} & \textbf{Obj. Det. Model} & \textbf{Runtime} \\
\hline
Energy cons.   & Jetson Orin Nano        & SSD v1          & TensorRT  \\ \hline
Inference time       & Raspberry Pi 5 + TPU    & SSD v1          & TFLite    \\ \hline
mAP (Group 1)        & Raspberry Pi 5 + TPU    & SSD v1          & TFLite    \\ \hline
mAP (Group 2)        & Raspberry Pi 5 + TPU    & SSD Lite        & TFLite    \\ \hline
mAP (Group 3)        & Jetson Orin Nano        & YOLOv8 Small    & TensorRT  \\ \hline
mAP (Group 4)        & Raspberry Pi 5 + AI HAT & YOLOv8 Small    & HEF       \\ \hline
mAP (Group 5)        & Raspberry Pi 5 + AI HAT & YOLOv8 Small    & HEF       \\ \hline

\end{tabular}
\vspace{-0.3cm}
\end{table}

In the experimental testbed, a subset of the profiled devices is deployed as the heterogeneous edge layer managed by the proposed load balancer. We use five physical devices as edge nodes, selected from the configurations in Table~\ref{tab:testbed_pairs}. These nodes host different device-model pairs and serve as the targets for task assignment decisions. A dedicated Raspberry Pi 4 is configured as the gateway device and runs the load balancing logic, including the implementation of the proposed load balancer and the communication with the edge nodes. Another Raspberry Pi 4 acts as the client device and emulates an IoT camera or sensing endpoint by sending inference requests to the gateway according to the workload scenarios in our experiments. The roles and hardware specifications of all devices in the testbed, including the client, gateway, and edge nodes, are summarised in Table~\ref{tab:testbed_config}. This setup reflects a realistic deployment in which resource constrained clients assign processing tasks to nearby heterogeneous edge nodes through an intermediate gateway.


\begin{table}[t]
\scriptsize
\centering
\caption{Device specifications for the experimental testbed.}
\label{tab:testbed_config}
\setlength{\tabcolsep}{3pt}
\begin{tabular}{|p{1.4cm}|p{2.9cm}|p{0.8cm}|p{0.8cm}|p{1.9cm}|}
\hline
\textbf{Role} & \textbf{Device Name} & \textbf{Proc.} & \textbf{Mem.} & \textbf{OS / SDK} \\
\hline
Client      & Raspberry Pi 4            & CPU      & 4 GB & Debian Bookworm \\ \hline
Gateway     & Raspberry Pi 4            & CPU      & 4 GB & Debian Bookworm \\ \hline
Edge node 1 & Raspberry Pi 5 + Coral TPU & TPU      & 4 GB & Debian Bookworm \\ \hline
Edge node 2 & Raspberry Pi 5 + Coral TPU & TPU      & 4 GB & Debian Bookworm \\ \hline
Edge node 3 & Raspberry Pi 5 + AI HAT    & NPU  & 4 GB & Debian Bookworm \\ \hline
Edge node 4 & Jetson Orin Nano           & GPU      & 8 GB & JetPack 5.1.3   \\ \hline
Edge node 5 & Jetson Orin Nano           & GPU      & 8 GB & JetPack 5.1.3   \\ \hline
\end{tabular}
\vspace{-0.3cm}
\end{table}

\subsection{Experimental Procedure}

To rigorously assess the effectiveness of the proposed load balancer, we compare it against a set of baseline load balancing strategies that represent commonly used policies in distributed systems. These baseline task assignment strategies are defined as follows.

\begin{itemize}
    \item \textbf{Round Robin (RR).} Distributes incoming image requests sequentially across all available edge nodes in a cyclic fashion.

    \item \textbf{Random (RND).} Assigns each request to an edge node chosen uniformly at random, independent of any system metrics or workload history.

    \item \textbf{Least Connection (LC).} Assigns each incoming request to the edge node with the smallest number of active requests or ongoing connections.
\end{itemize}

In addition to these generic baselines, we include three single objective policies that approximate best achievable performance for each metric of interest.

\begin{itemize}
    \item \textbf{Lowest Energy (LE).} Assigns every image to the device-model pair with the lowest profiled energy consumption per request. In practice, this corresponds to a single fixed configuration that is always selected, providing an approximate lower bound on energy consumption while ignoring latency and accuracy.

    \item \textbf{Lowest Time (LT).} Selects the edge node that minimises an estimate of expected delay, obtained by multiplying the current number of requests in the queue at that node by the offline profiled inference time of the corresponding device and model pair. This policy focuses on latency reduction and provides an approximate lower bound on response time, without explicitly considering energy or accuracy.

    \item \textbf{Highest Accuracy (HA).} Always selects the device-model pair with the highest profiled mAP. As with LE, this yields a single fixed configuration that is used for all requests and provides an approximate upper bound on achievable accuracy, regardless of energy consumption or latency.
\end{itemize}

To capture different application requirements, we vary the trade off parameter $\gamma$ in the weighted sum scoring function, which controls the relative importance of latency and energy in the decision process. Small values of $\gamma$ favour energy efficiency, larger values emphasise latency, and intermediate settings provide a balanced operating point.

To model temporal variations in workload intensity, we generate different levels of concurrency using Locust.\footnote{https://locust.io/} In Locust, each user process issues a single image request and then waits until a response is received before sending the next one. Thus, the number of concurrent users directly corresponds to the number of requests that can be in progress at the same time. In our experiments, we vary the concurrency level by adjusting the number of users to $\{1, 3, 5, ..., 15\}$. This configuration allows us to evaluate the behaviour of the proposed load balancer and the baselines under both low-load and progressively higher-load scenarios, where multiple requests arrive simultaneously and must be assigned to edge nodes in parallel.

The performance of each load balancing strategy is evaluated using the following primary metrics:

\textbf{Energy consumption efficiency.} Reported as the average energy consumed per request, excluding the base idle energy of each device. Energy measurements are obtained from UM25C USB power meters attached to every edge node and expressed in milliwatt hours, so that the metric captures only the energy attributable to processing workload requests.

\textbf{Latency (average response time):} Reported in milliseconds as the average response time per request, as provided by Locust. For each experiment, Locust measures the time from when a request is issued by the client until the corresponding response is received, and computes the mean across all completed requests.

\textbf{Latency p90:} Reported in milliseconds as the 90th percentile of the response time distribution, also obtained from Locust’s statistics. This value indicates the response time below which 90\% of all requests complete and is used to characterise tail latency for latency-sensitive applications.

\textbf{Throughput:} Measured in requests per second, using the aggregate \emph{requests/s} value reported by Locust over the duration of each experiment. This metric reflects the overall processing rate achieved under a given concurrency level and load balancing strategy.

\textbf{Accuracy:} Assessed using mAP, computed with the FiftyOne tool,\footnote{https://docs.voxel51.com/} which compares the detection outputs against the reference annotations and provides robust metrics for evaluating object detection performance.

To enhance statistical reliability, each experimental configuration was executed 3 times under identical settings; we report the mean across runs and characterize run-to-run variability using error bars.

\subsection{Experimental Results}
\subsubsection{Proposed Load Balancer against Baseline Strategies}
\begin{figure*}[htbp]
    \centering
    \subfigure[Average Response Time]{
        \includegraphics[width=0.31\textwidth]{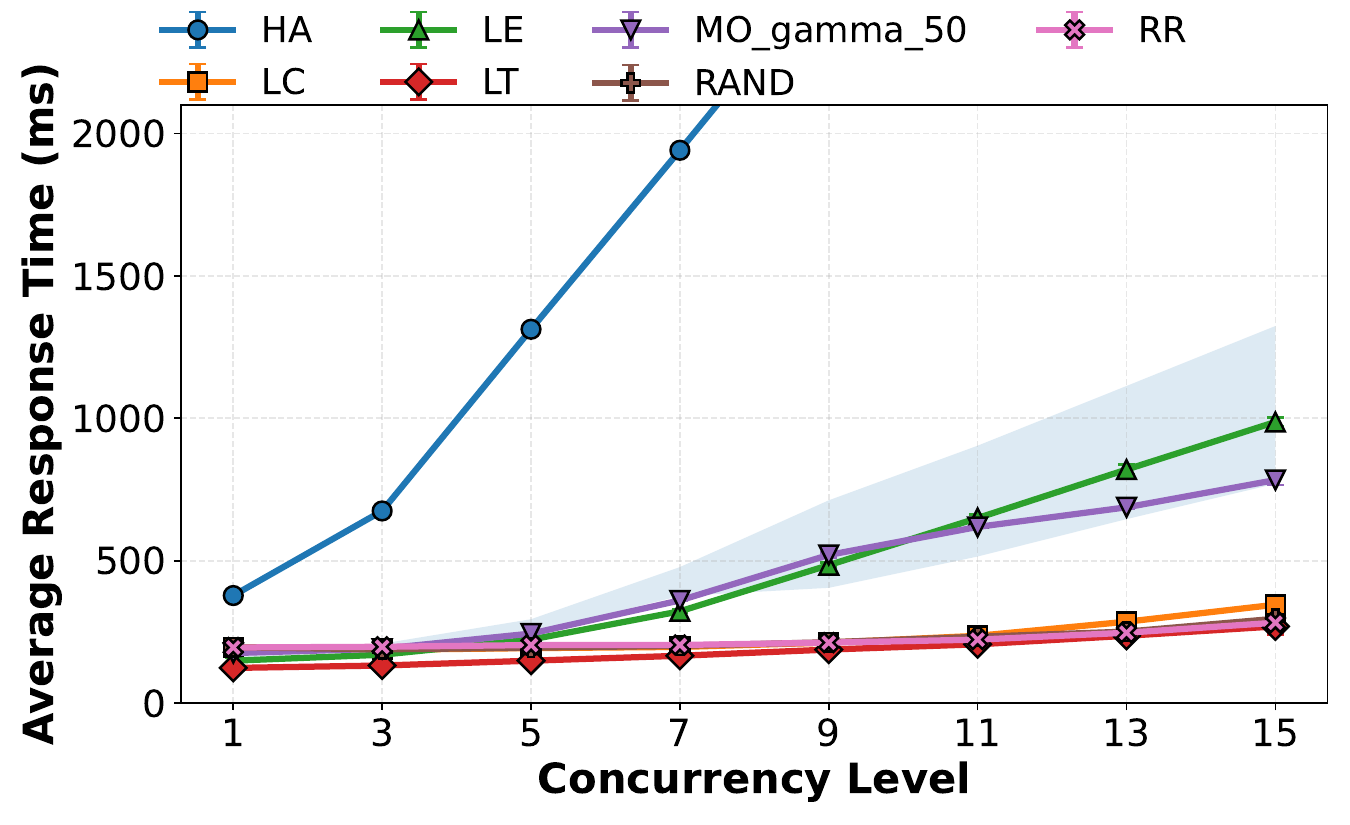}
    }
    \subfigure[Latency p90]{
        \includegraphics[width=0.31\textwidth]{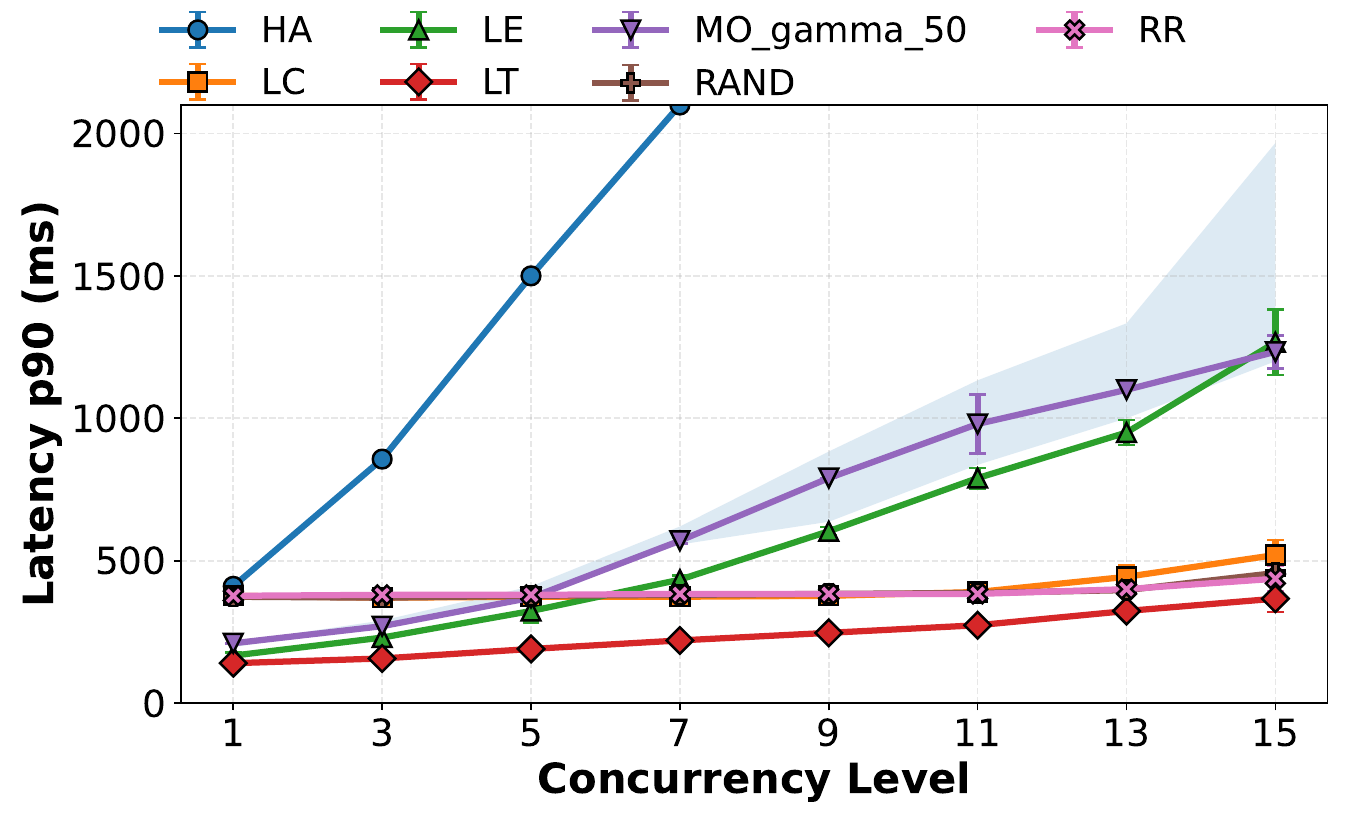}
    }
    \subfigure[CDF]{
        \includegraphics[width=0.31\textwidth]{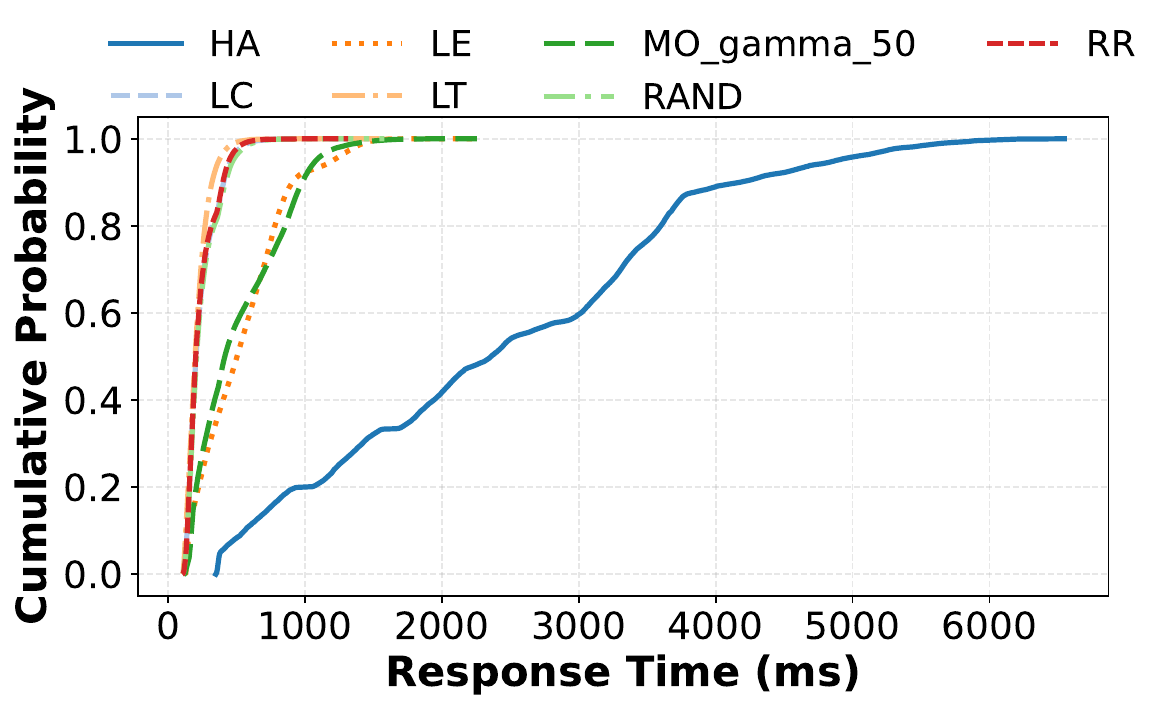}
    }
    \subfigure[Throughput]{
        \includegraphics[width=0.31\textwidth]{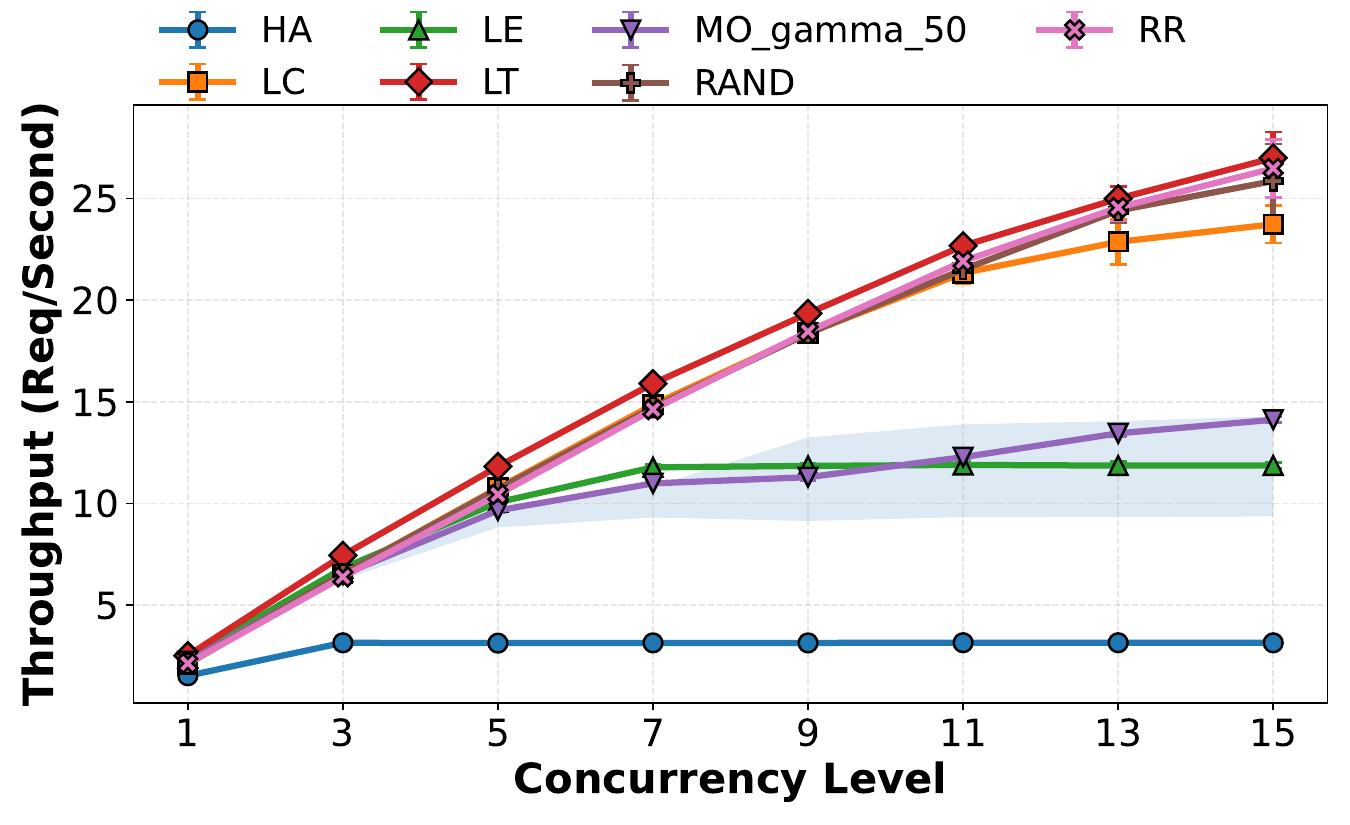}
    }
    \subfigure[Energy Consumption]{
        \includegraphics[width=0.31\textwidth]{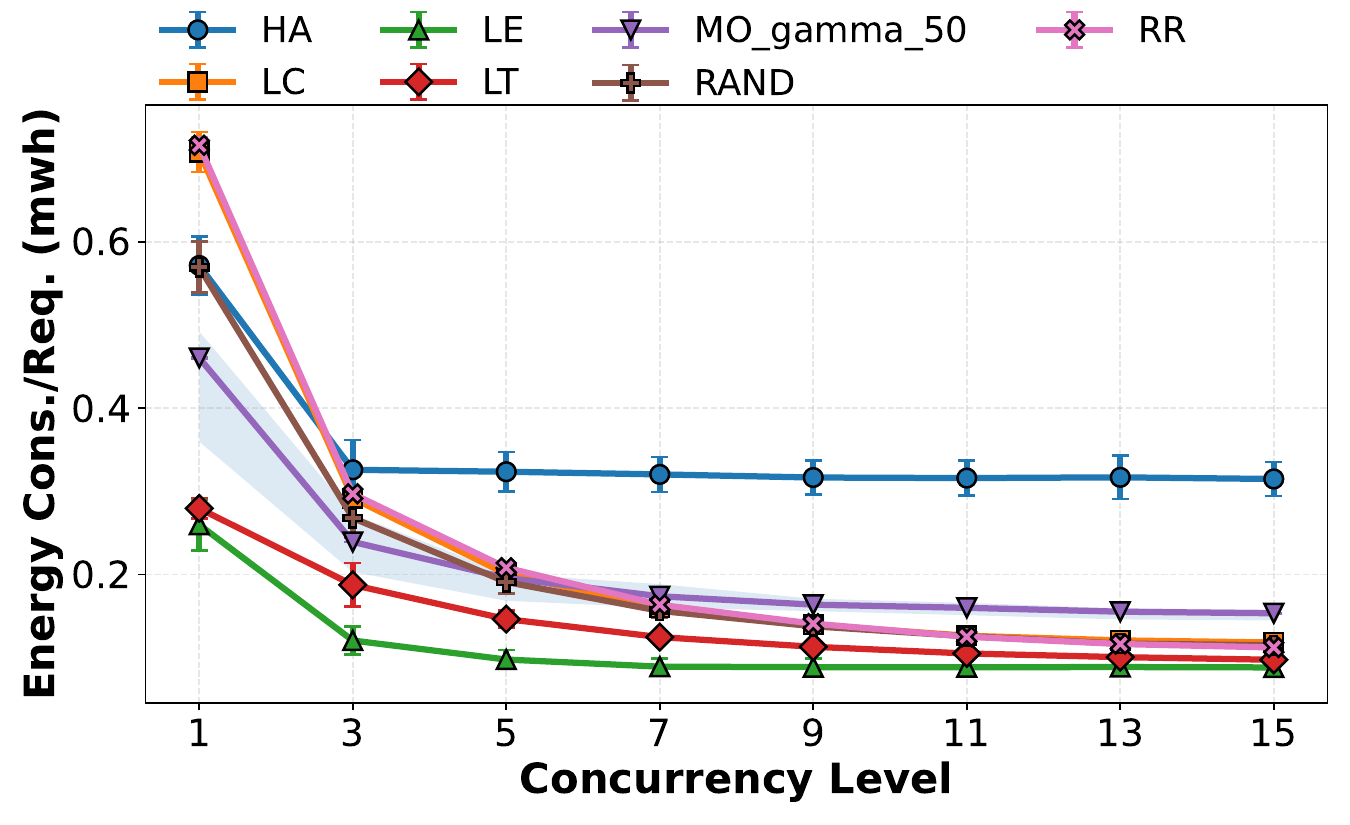}
    }
    \subfigure[Accuracy (mAP)]{
        \includegraphics[width=0.31\textwidth]{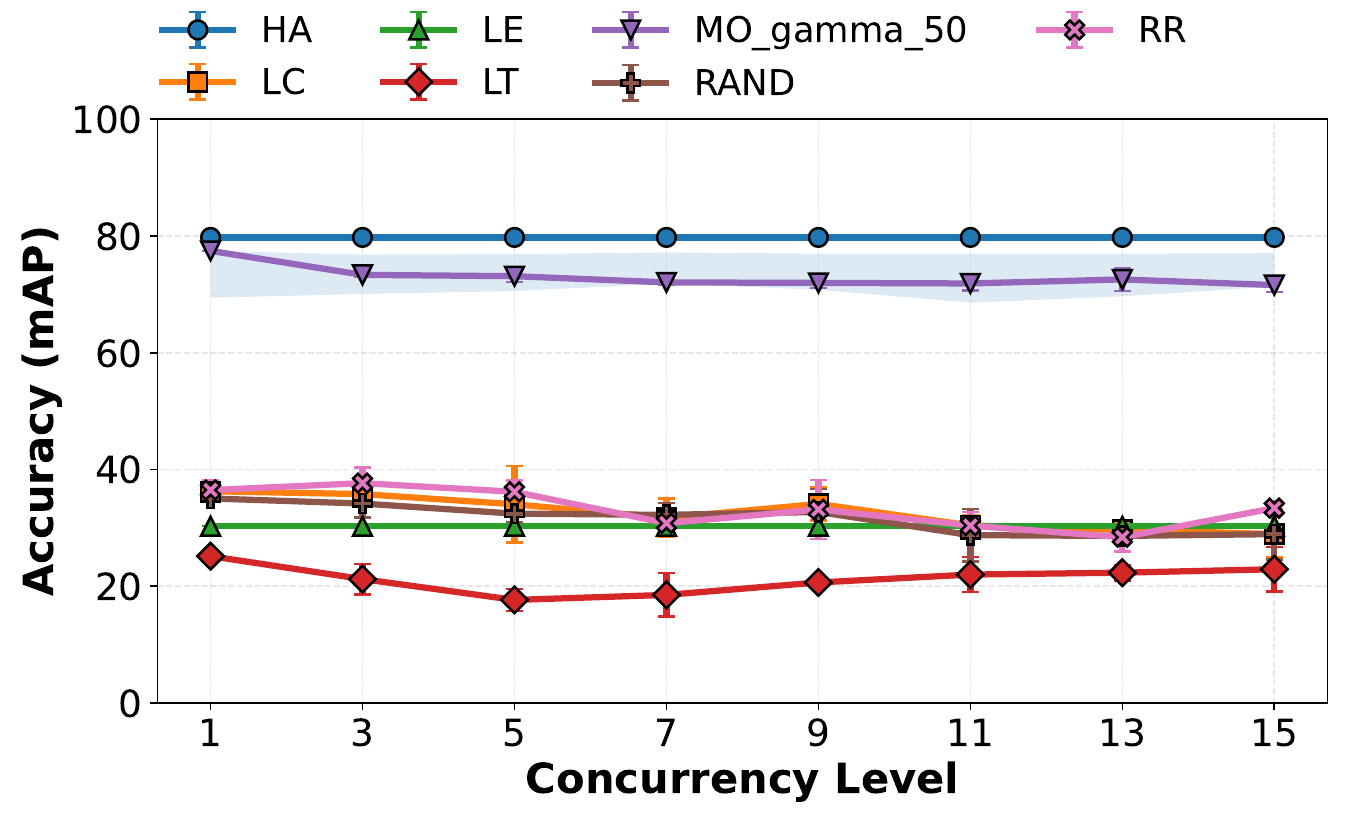}
    }
   
    \caption{Comparison of load balancing policies across different metrics under varying request concurrency. The shaded region around MO\_gamma\_50 indicates the range between MO\_gamma\_0 and MO\_gamma\_1.}
    \vspace{-0.3 cm}
    \label{fig:results_proposed and basedlines}
\end{figure*}

We first compare the proposed load balancer against the baseline strategies across different concurrency levels. Figure~\ref{fig:results_proposed and basedlines} reports the main performance metrics as a function of the number of concurrent users, where the curves correspond to the multi objective configuration with $\gamma = 0.50$ (denoted MO\_gamma\_50) and the baseline methods RR, RND, LC, LE, LT, and HA.

The behaviour of average response time in Fig.~\ref{fig:results_proposed and basedlines}(a) illustrates the overall latency trends as the system becomes more heavily loaded. For every method, response time increases with the number of concurrent users, but with different magnitudes. As expected, the latency oriented baseline LT achieves the smallest response times across all loads, since it explicitly optimises delay and ignores energy and accuracy. MO\_gamma\_50 has higher latency than LT, especially at the highest load, but remains substantially faster than the accuracy oriented HA and comparable to the generic load spreading strategies RR, RND, and LC. At 15 concurrent users, MO\_gamma\_50 reaches an average response time of about 790 ms, compared with 260 ms for LT, 977 ms for LE, and 4436 ms for HA, so it incurs roughly three times the latency of LT at this highest load while still reducing latency by about 20 percent relative to LE and by more than 80 percent relative to HA. This shows that MO\_gamma\_50 accepts a moderate increase in delay compared with the latency lower bound in order to achieve much better energy and accuracy behaviour than the extreme baselines.

Tail latency, measured by the p90 metric in Fig.~\ref{fig:results_proposed and basedlines}(b), exhibits a similar pattern. As concurrency increases, the ninetieth percentile of response time grows for all methods, but LT maintains the smallest p90 values and thus the tightest control over extreme delays. HA displays the worst tail behaviour, with p90 increasing sharply due to its systematic selection of the slowest but most accurate configuration. MO\_gamma\_50 and LE form an intermediate group that remains substantially below HA in terms of p90 but above LT and the simple load spreading baselines at medium and high concurrency levels. In this regime, MO\_gamma\_50 keeps tail latency within an acceptable range while enforcing the accuracy tolerance and accounting for energy, rather than optimising delay alone.

The cumulative distribution function of response time in Fig.~\ref{fig:results_proposed and basedlines}(c) further corroborates these latency trends. The curve for LT lies furthest to the left, confirming it as the latency lower bound, while HA has the rightmost distribution with a large fraction of requests experiencing multi second delays. MO\_gamma\_50 tracks close to LT and the generic load spreading baselines RR, RND, and LC over most of the distribution, yet remains clearly to the left of both LE and HA. This indicates that MO\_gamma\_50 completes a larger fraction of requests within a given latency budget than the energy oriented and accuracy oriented single objective baselines, while still respecting the accuracy tolerance and considering energy in the decision process.

Throughput, measured in requests per second and reported in Fig.~\ref{fig:results_proposed and basedlines}(d), highlights how efficiently each strategy utilises the available edge resources under increasing load. LT, RR, RND, and LC reach the highest throughput values, since they keep all nodes busy without restricting themselves to a subset of configurations. MO\_gamma\_50 operates at a moderate throughput level that is lower than these purely throughput oriented strategies but substantially higher than HA and slightly above LE at most concurrency levels. At 15 concurrent users, MO\_gamma\_50 achieves approximately 15 requests per second, compared with around 25 requests per second for LT, RR, RND, and LC, about 12 requests per second for LE, and roughly 3 requests per second for HA. Thus, the proposed method trades some throughput relative to the latency lower bound while still delivering around five times higher throughput than the accuracy upper bound.

Energy consumption per request provides the complementary view of efficiency and is shown in Fig.~\ref{fig:results_proposed and basedlines}(e). The average energy per request, excluding base idle energy, decreases as concurrency increases and then stabilises once devices become well utilised. Among all strategies, LE achieves the lowest energy consumption across all loads, since it always selects the most energy efficient device and model pair, and LT is very close to LE and effectively represents the second lowest energy profile. MO\_gamma\_50 consumes more energy than LE and LT but remains similar to or lower than RR, RND, and LC and clearly below HA at all concurrency levels. At 15 concurrent users, LE and LT both consume around 0.10 mWh per request, MO\_gamma\_50 about 0.15 mWh, and HA about 0.30 mWh, so MO\_gamma\_50 uses roughly 33 percent more energy per request than the energy lower bound but still halves the energy cost relative to the accuracy upper bound.

The accuracy results, expressed in mAP and reported in Fig.~\ref{fig:results_proposed and basedlines}(f), show why this combined behaviour is attractive. At low load with one concurrent user, MO\_gamma\_50 achieves an mAP of around 78 and is very close to HA, which stays near 79 across all loads. As concurrency increases, the accuracy of MO\_gamma\_50 decreases slightly and stabilises at approximately 71 mAP, about 8 points, around 10 percent, below HA, but still substantially higher than all other baselines. RR and LC remain around the mid 30 mAP range, LE and RND around 30 mAP, and LT between 20 and 25 mAP. At high concurrency, MO\_gamma\_50 therefore achieves more than twice the accuracy of LT and more than double that of LE in relative terms, while keeping energy and latency at levels that are much closer to the lower bounds than to the accuracy upper bound.

\medskip
\noindent\textbf{Insight.} Across all metrics, MO\_gamma\_50 achieves a balanced operating point. It does not minimise latency or energy in isolation, but offers a controlled trade off in which average delay remains within a moderate factor of the latency lower bound, energy per request stays close to the most efficient configurations and far below the accuracy upper bound, and mAP remains within roughly 10 percent of the best achievable accuracy while more than doubling the accuracy of energy and latency oriented baselines. This demonstrates that enforcing an accuracy tolerance and optimising jointly over latency and energy yields a load balancing policy that is competitive across all metrics rather than optimal in only one.

\subsection{Results Across Different $\gamma$ Values}

\begin{figure*}[htbp]
    \centering
    \subfigure[Average Response Time]{
        \includegraphics[width=0.31\textwidth]{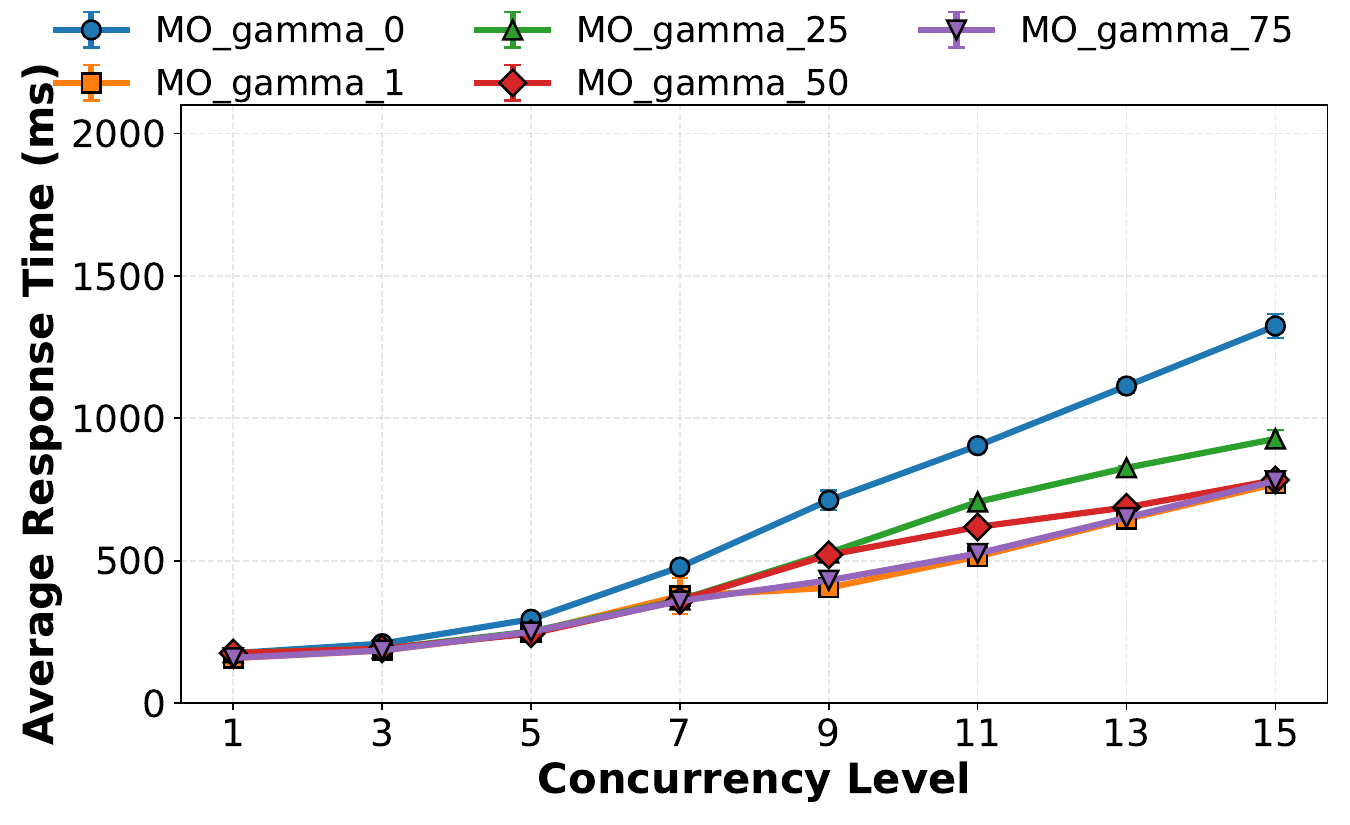}
    }
    \subfigure[Latency p90]{
        \includegraphics[width=0.31\textwidth]{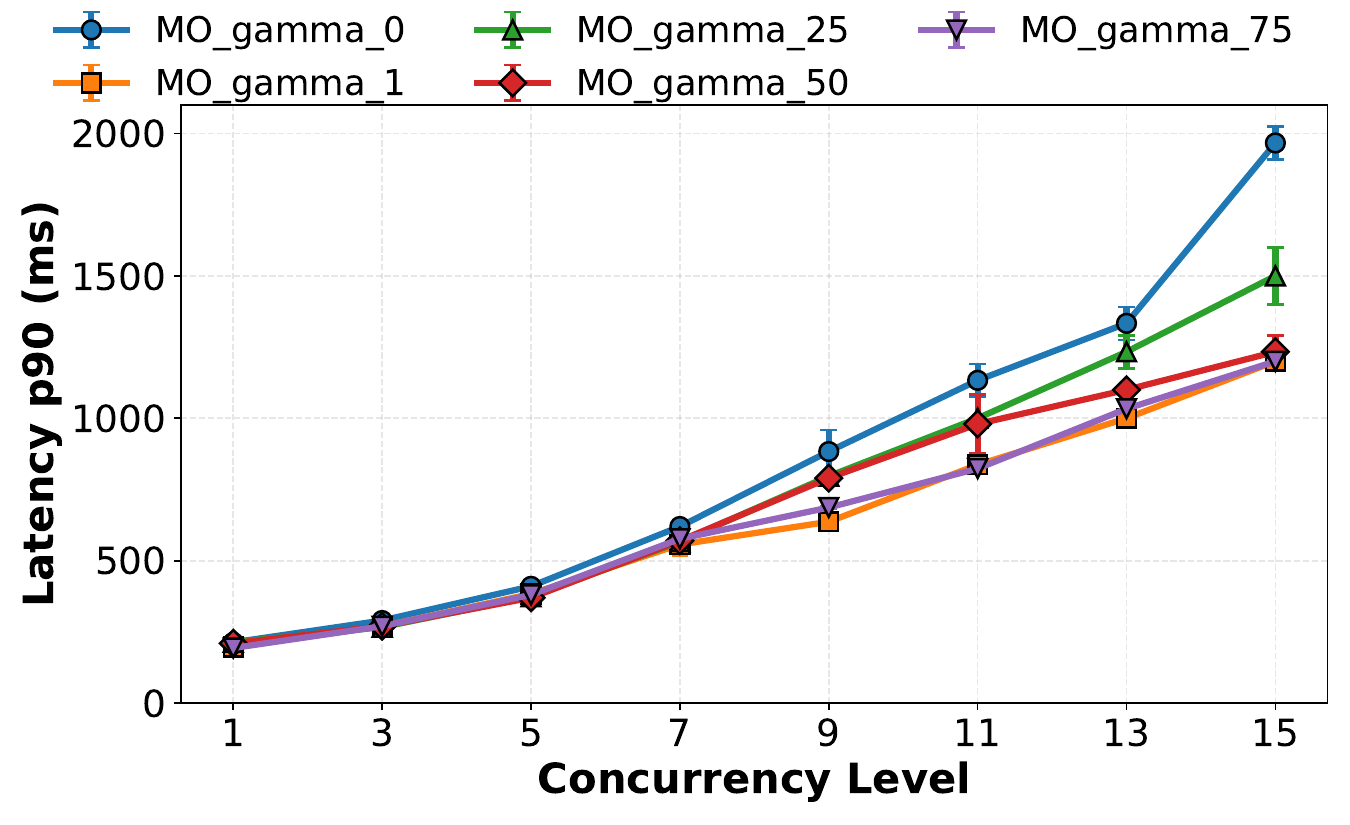}
    }
    \subfigure[Throughput]{
        \includegraphics[width=0.31\textwidth]{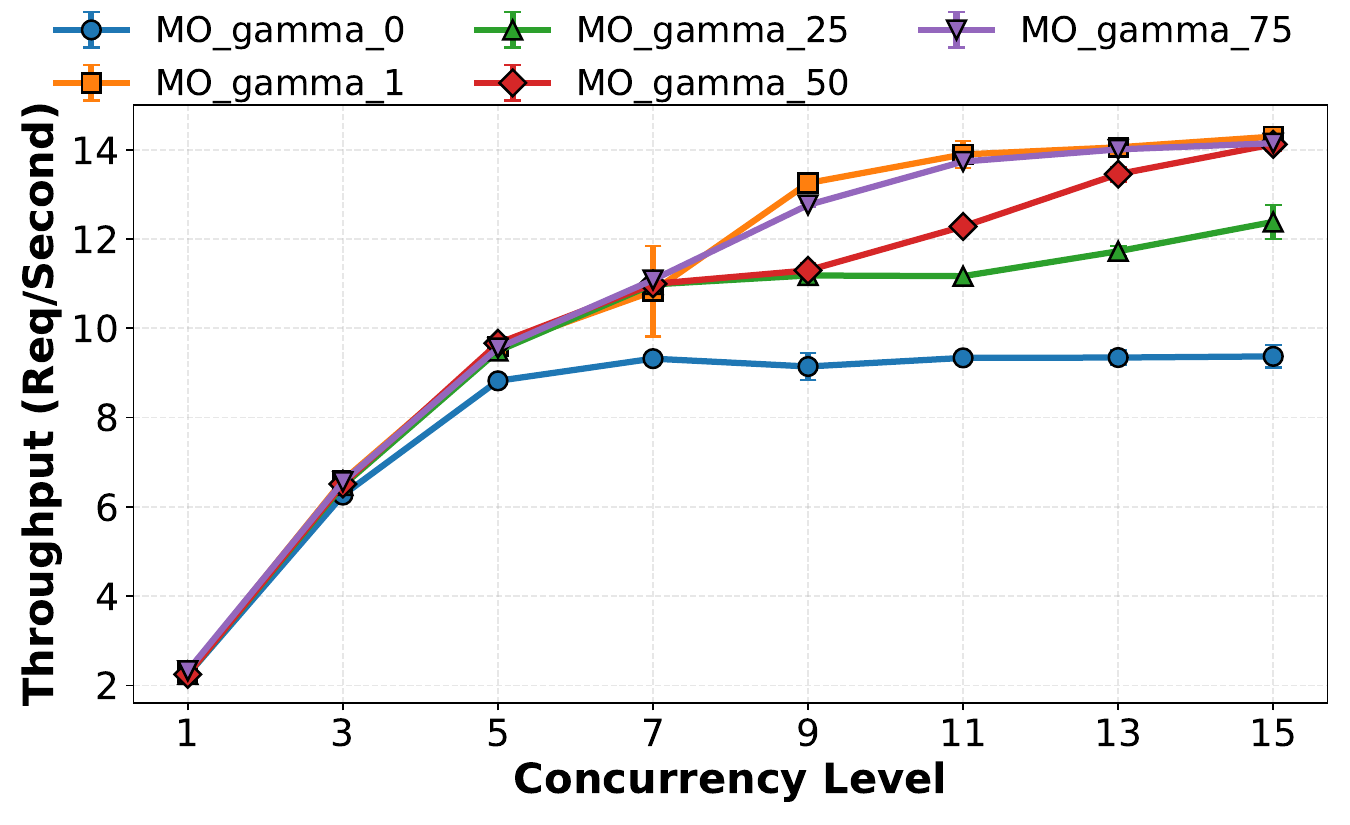}
    }
    \subfigure[Energy Consumption]{
        \includegraphics[width=0.31\textwidth]{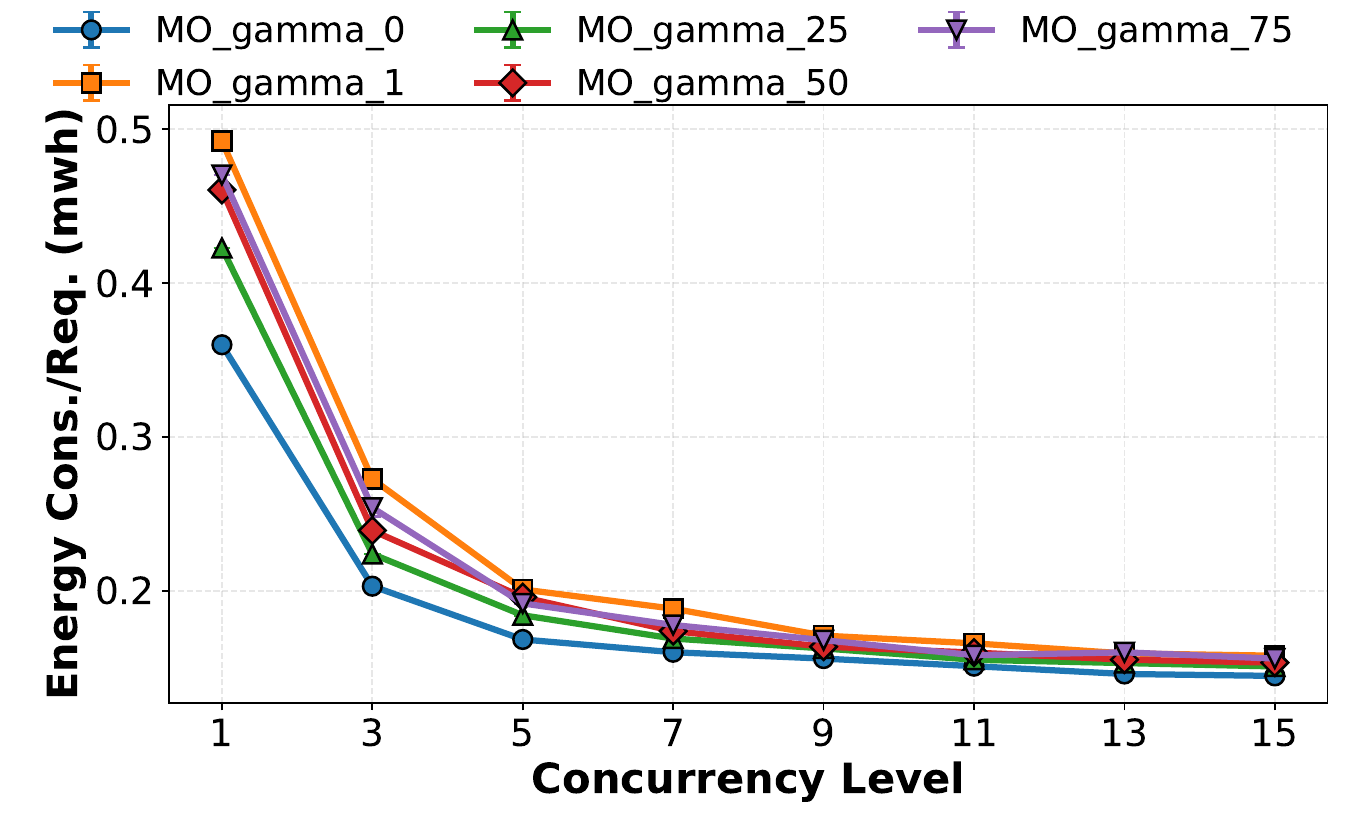}
    }
    \subfigure[Accuracy]{
        \includegraphics[width=0.31\textwidth]{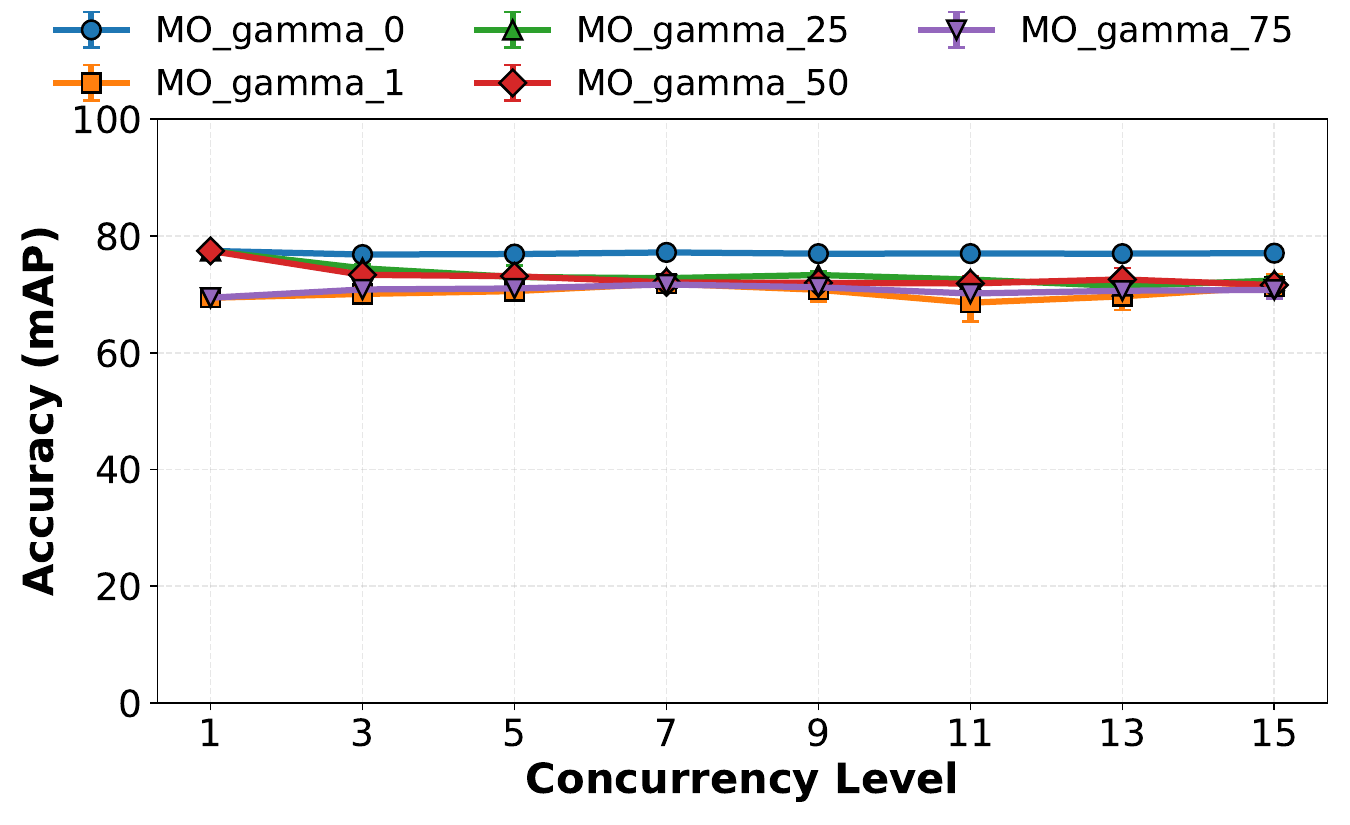}
    }

    \caption{Comparison of proposed load balancer with changing gamma value across different metrics under varying request concurrency.}
    \vspace{-0.3 cm}
    \label{fig:results_gamma}
\end{figure*}

To analyse how the value of parameter $\gamma$ influences the behaviour of the proposed multi-objective load balancer, we evaluate five configurations, denoted MO\_gamma\_0, MO\_gamma\_25, MO\_gamma\_50, MO\_gamma\_75, and MO\_gamma\_1, corresponding to $\gamma \in \{0, 0.25, 0.50, 0.75, 1\}$. Recall that smaller values of $\gamma$ place more emphasis on energy consumption, whereas larger values prioritise latency.

The evolution of average response time with concurrency levels is shown in Fig.~\ref{fig:results_gamma}(a). As concurrency increases, all configurations experience higher latency, but the magnitude of this increase depends strongly on $\gamma$. The energy-oriented variant MO\_gamma\_0 consistently yields the largest average response time across all loads, reaching roughly 1.3--1.4 seconds at 15 concurrent users. As $\gamma$ increases, the curves progressively shift downward. MO\_gamma\_25 reduces average latency compared with MO\_gamma\_0, and MO\_gamma\_50, MO\_gamma\_75, and MO\_gamma\_1 achieve the lowest response times, remaining tightly clustered. At 15 concurrent users, MO\_gamma\_1 exhibits an average response time of around 750~ms, whereas MO\_gamma\_0 is close to 1350~ms, corresponding to an improvement of approximately 40--45\% when moving from a purely energy-oriented to a strongly latency-oriented configuration. MO\_gamma\_50 and MO\_gamma\_75 lie close to MO\_gamma\_1, indicating that moderate values of $\gamma$ already capture most of the latency benefits.

A similar monotonic effect of $\gamma$ is observed for the tail of the latency distribution. Fig.~\ref{fig:results_gamma}(b) reports the 90th percentile of response time (p90) for the same configurations. For low concurrency levels, the differences between the five curves are relatively small, but as the load increases the separation becomes more pronounced. MO\_gamma\_0 again exhibits the worst tail latency, with p90 approaching 1.7--1.8~s at 15 concurrent users, while the latency-oriented variants MO\_gamma\_75 and MO\_gamma\_1 keep p90 close to 1.1--1.2~s at the same load. This represents a reduction of roughly one third in p90 when shifting from $\gamma = 0$ to $\gamma \approx 1$. MO\_gamma\_25 and MO\_gamma\_50 trace intermediate trajectories and demonstrate that gradual increases in $\gamma$ yield progressive improvements in tail latency.

Throughput results, shown in Fig.~\ref{fig:results_gamma}(c), further illustrate the impact of $\gamma$ on the utilisation of the edge resources. All configurations start from similar throughput at low concurrency, but diverge as the number of concurrent users grows. MO\_gamma\_0 saturates at around 9--10~requests/s, whereas MO\_gamma\_25 reaches approximately 12--13~requests/s. The more latency-oriented variants MO\_gamma\_50, MO\_gamma\_75, and MO\_gamma\_1 achieve the highest throughput, approaching 14--15~requests/s at 15 concurrent users. Comparing the extremes at this highest load, MO\_gamma\_1 provides roughly 50\% higher throughput than MO\_gamma\_0, again highlighting the effect of prioritising faster configurations in the scoring function.

The energy consumption per request, excluding base idle energy, is reported in Fig.~\ref{fig:results_gamma}(d). For all values of $\gamma$, the energy per request decreases between one and five concurrent users, reflecting improved utilisation, and then stabilises for higher loads. In contrast to the latency and throughput metrics, the ordering with respect to $\gamma$ is reversed. MO\_gamma\_0 consistently achieves the lowest energy consumption, with values around 0.14--0.15~mWh per request at 15 concurrent users, while MO\_gamma\_1 attains the highest energy consumption, close to 0.17~mWh per request. Intermediate settings such as MO\_gamma\_25 and MO\_gamma\_50 lie between these extremes; for example, MO\_gamma\_50 consumes roughly 0.16~mWh per request at 15 concurrent users. Overall, moving from $\gamma = 0$ to $\gamma = 1$ increases energy consumption per request by approximately 15--20\%, while MO\_gamma\_50 incurs around 10\% higher energy than MO\_gamma\_0 at high concurrency.

Accuracy, measured in mAP, is much less sensitive to the choice of $\gamma$. Fig.~\ref{fig:results_gamma}(e) shows that all configurations maintain relatively high and stable mAP values across the evaluated concurrency levels. MO\_gamma\_0 achieves the highest accuracy, typically around 76--78~mAP, and exhibits a slight upward trend as concurrency increases. The remaining configurations---MO\_gamma\_25, MO\_gamma\_50, MO\_gamma\_75, and MO\_gamma\_1---cluster around 70--72~mAP, with only minor fluctuations over the range of concurrency levels. At 15 concurrent users, the gap between MO\_gamma\_0 and the other variants is on the order of 6--8~mAP points, corresponding to roughly an 8--10\% difference. This indicates that, while favouring energy (small $\gamma$) can yield modest gains in accuracy, the impact of $\gamma$ on mAP is substantially smaller than its impact on latency, throughput, or energy consumption.

\medskip
\noindent\textbf{Insight.} The results show that the parameter $\gamma$ provides an effective mechanism for tuning the trade off between latency, throughput, and energy in the proposed load balancer, with only modest impact on accuracy. Among the evaluated settings, MO\_gamma\_50 emerges as a balanced configuration that captures most of the latency and throughput benefits of higher $\gamma$ values while keeping the increase in energy consumption moderate and maintaining accuracy within a narrow and acceptable range.

\section{Discussions}

The experimental results demonstrate that the proposed multi-objective load balancer is able to exploit device and model heterogeneity more effectively than static or single-objective baselines. By combining an accuracy-aware filtering stage with a weighted-sum scoring function over latency and energy, the load balancer maintains high mAP that is close to the accuracy-oriented baseline, while substantially reducing energy consumption compared with the accuracy-oriented and keeping average response time and tail latency within a moderate factor of the best latency-oriented strategy. The trade-off parameter $\gamma$ further provides a practical control knob that allows operators to tune the system between energy-saving and latency-driven regimes without redesigning the algorithm, and the evaluation across multiple concurrency levels shows that these properties hold under both light and heavy load.

At the same time, the current design relies on offline profiling of device and model pairs and does not adapt the performance data at run time, which may reduce robustness if workloads or hardware characteristics drift over time. Additionally, the testbed is limited to a single application domain and a modest number of edge nodes, and all load balancing decisions are made at a central gateway. For small-scale deployments similar to our setup, this centralised design is practical and does not introduce a noticeable bottleneck; however, in larger edge clusters with many nodes and high request rates, a single gateway may become a scalability constraint, and distributed or hierarchical load balancing architectures could be more appropriate.

\section{Related work}
\label{sec:related_work}

To our knowledge, no existing work addresses the same multi objective load balancing problem for heterogeneous edge based object detection. We therefore first review task offloading and scheduling, then load balancing and resource allocation in heterogeneous edge environments, and finally deep neural network (DNN) based object detection at the edge.

\subsection{Edge Computing and Resource Management}
A substantial body of work investigates resource management in edge and mobile edge computing, focusing on task offloading, scheduling, and autoscaling. For task offloading, several studies formulate multi objective optimisation problems that jointly minimise latency and energy across heterogeneous edge servers, for example computation sharing and cooperative mobile edge schemes that coordinate offloading and power control~\cite{bozorgchenani2021multi,fan2025latency}. At the scheduling level, other work allocates tasks over heterogeneous compute resources under delay and resource constraints, including bi objective schedulers that trade communication cost against computation delay and frameworks for serving heterogeneous machine learning models on shared accelerators~\cite{hmimz2021biobjective,choi2022gpulet}. Complementary research addresses autoscaling of edge hosted services, where predictive and distributed autoscalers for Kubernetes based edge clusters dynamically adjust resource allocations to balance utilisation and quality of service~\cite{ju2022proactive,spatharakis2022autoscaling}. While these approaches demonstrate the effectiveness of optimisation driven and learning based resource management, they mainly optimise latency, energy, and utilisation at the platform level, do not explicitly integrate object detection accuracy and device specific profiling, and do not directly address load balancing when workloads and scene complexity vary over time, which is the focus of our work.

\subsection{Load Balancing and Resource Allocation at the Edge}

A large body of work has investigated load balancing and resource allocation mechanisms for edge and MEC platforms. Aslanpour et al.~\cite{aslanpour2024load} propose load balancing policies for heterogeneous serverless edge infrastructures that account for multiple performance indicators, including latency, energy efficiency, and AI-related metrics, using empirical profiling to tune the decision logic. Avgeris et al.~\cite{avgeris2022enerdge} introduce a distributed energy-aware resource allocation framework that combines load redirection with Markov Random Fields to redistribute excess load while reducing energy consumption in IoT edge deployments. Chen et al.~\cite{chen2021enhancing} design a load balancing algorithm for MEC that leverages user-load prediction to estimate server utilisation and improve task offloading decisions in ultra-dense networks. These studies typically focus on generic compute or MEC workloads and rarely integrate application-level accuracy into the decision process. Moreover, most of these approaches assume relatively coarse-grained characterisations of node performance (for example, aggregate CPU capacity, queue length, or average power consumption), whereas our formulation uses fine-grained offline profiling of specific device--model pairs, including accuracy, latency, and energy as functions of scene complexity.

\subsection{Deep Neural Network Inference and Object Detection at the Edge}

In parallel, there has been extensive interest in optimising deep neural network (DNN) inference for resource constrained edge devices. Ngo et al.\ survey edge intelligence techniques and identify key trade offs between model complexity, latency, and energy consumption across diverse hardware platforms and optimisation methods~\cite{ngo2025edgeintelligence}, while Yan et al.\ and Rachuri et al.\ propose runtime mechanisms that adapt execution to workload conditions to reduce energy usage and tail latency~\cite{yan2023polythrottle,rachuri2024ecoedgeinfer}. Complementary work focuses on model partitioning and collaborative execution, including split computing between IoT devices and nearby edge servers~\cite{karjee2022split}, distributed inference of directed acyclic graph structured networks across heterogeneous edge resources~\cite{hu2022edgeflow}, and service function chaining for low latency object detection in softwarised networks~\cite{xiang2021yolocompute}. These contributions address important aspects of DNN inference at the edge but generally treat accuracy as a fixed property of the chosen model rather than an explicit constraint in the decision process and often target a single device or a homogeneous server cluster. In contrast, our work focuses on object detection in smart city scenarios using a heterogeneous testbed and formulates load balancing as a multi objective task allocation problem across multiple device-model pairs.

\section{Conclusion and Future Works}
\label{sec:conclusions}

This paper has presented a multi-objective load balancer for heterogeneous edge computing, designed for object detection workloads in smart city scenarios. The proposed approach combines an accuracy-aware filtering stage, which excludes device-model pairs that do not satisfy a given accuracy threshold, with a weighted-sum scoring function that jointly considers expected latency and energy consumption.

The experimental results show that the proposed load balancer can navigate the trade-offs between latency, energy, throughput, and accuracy more effectively than the baselines. Compared with the accuracy-oriented strategy, the proposed configuration with $\gamma = 0.50$ reduces average response time by more than 80\% and halves the energy consumption per request at high concurrency, while remaining within roughly 10\% of its mAP. The analysis of different $\gamma$ values further demonstrates that the same mechanism can be tuned to favour either energy or latency, with only modest impact on accuracy, providing a practical control knob for operators to adapt the behaviour of the system to application-specific requirements.

\textbf{Future Work:} To enhance the robustness of the multi objective load balancer, future work will extend the load balancing strategy to incorporate online measurements or learning based updates to performance estimates and to explore distributed or hierarchical designs for larger edge deployments.



\bibliographystyle{ieeetr}
\bibliography{load_balancing_ref}

@article{daghash2024,
  title={Benchmarking Deep Learning Models for Object Detection on Edge Computing Devices},
  author={Alqahtani, Daghash K and Cheema, Aamir and Toosi, Adel N},
  journal={arXiv preprint arXiv:2409.16808},
  year={2024}
}

@article{alqahtani2025,
  title={ECORE: Energy-Conscious Optimized Routing for Deep Learning Models at the Edge},
  author={Alqahtani, Daghash K and Rodriguez, Maria A and Cheema, Muhammad Aamir and Rezatofighi, Hamid and Toosi, Adel N},
  journal={arXiv preprint arXiv:2507.06011},
  year={2025}
}

@InProceedings{Redmon2016,
  author    = {Redmon, Joseph and Divvala, Santosh and Girshick, Ross and Farhadi, Ali},
  booktitle = {Proceedings of the IEEE conference on computer vision and pattern recognition},
  title     = {You only look once: Unified, real-time object detection},
  year      = {2016},
  pages     = {779--788},
}

@InProceedings{Liu2016,
  author       = {Liu, Wei and Anguelov, Dragomir and Erhan, Dumitru and Szegedy, Christian and Reed, Scott and Fu, Cheng-Yang and Berg, Alexander C},
  booktitle    = {Computer Vision--ECCV 2016: 14th European Conference, Amsterdam, The Netherlands, October 11--14, 2016, Proceedings, Part I 14},
  title        = {Ssd: Single shot multibox detector},
  year         = {2016},
  organization = {Springer},
  pages        = {21--37},
}

@InProceedings{Tan2020,
  author    = {Tan, Mingxing and Pang, Ruoming and Le, Quoc V},
  booktitle = {Proceedings of the IEEE/CVF conference on computer vision and pattern recognition},
  title     = {Efficientdet: Scalable and efficient object detection},
  year      = {2020},
  pages     = {10781--10790},
}

@software{ultralytics,
  author = {Glenn Jocher and Ayush Chaurasia and Jing Qiu},
  title = {Ultralytics YOLOv8},
  version = {8.0.0},
  year = {2023},
  url = {https://github.com/ultralytics/ultralytics},
  orcid = {0000-0001-5950-6979, 0000-0002-7603-6750, 0000-0003-3783-7069},
  license = {AGPL-3.0}
}

@misc{iotanalytics2025devices,
  author       = {{IoT Analytics}},
  title        = {Number of Connected {IoT} Devices},
  url = {https://iot-analytics.com/number-connected-iot-devices/},
  note         = {Accessed: 10~Dec.~2025},
  year         = {2025}
}

@article{aslanpour2024load,
  title={Load balancing for heterogeneous serverless edge computing: A performance-driven and empirical approach},
  author={Aslanpour, Mohammad Sadegh and Toosi, Adel N and Cheema, Muhammad Aamir and Chhetri, Mohan Baruwal and Salehi, Mohsen Amini},
  journal={Future generation computer systems},
  volume={154},
  pages={266--280},
  year={2024},
  publisher={Elsevier}
}

@article{avgeris2022enerdge,
  title={ENERDGE: Distributed energy-aware resource allocation at the edge},
  author={Avgeris, Marios and Spatharakis, Dimitrios and Dechouniotis, Dimitrios and Leivadeas, Aris and Karyotis, Vasileios and Papavassiliou, Symeon},
  journal={Sensors},
  volume={22},
  number={2},
  pages={660},
  year={2022},
  publisher={MDPI}
}

@article{chen2021enhancing,
  title={Enhancing mobile edge computing with efficient load balancing using load estimation in ultra-dense network},
  author={Chen, Wen and Zhu, Yongqi and Liu, Jiawei and Chen, Yuhu},
  journal={Sensors},
  volume={21},
  number={9},
  pages={3135},
  year={2021},
  publisher={MDPI}
}

@inproceedings{rachuri2024ecoedgeinfer,
  title={EcoEdgeInfer: Dynamically optimizing latency and sustainability for inference on edge devices},
  author={Rachuri, Sri Pramodh and Shaik, Nazeer and Choksi, Mehul and Gandhi, Anshul},
  booktitle={2024 IEEE/ACM Symposium on Edge Computing (SEC)},
  pages={191--205},
  year={2024},
  organization={IEEE}
}

@article{ngo2025edgeintelligence,
  title={Edge intelligence: A review of deep neural network inference in resource-limited environments},
  author={Ngo, Dat and Park, Hyun-Cheol and Kang, Bongsoon},
  journal={Electronics},
  volume={14},
  number={12},
  pages={2495},
  year={2025},
  publisher={MDPI}
}

@article{yan2023polythrottle,
  title={PolyThrottle: Energy-efficient Neural Network Inference on Edge Devices},
  author={Yan, Minghao and Wang, Hongyi and Venkataraman, Shivaram},
  journal={arXiv preprint arXiv:2310.19991},
  year={2023}
}

@article{karjee2022split,
  title={Split computing: DNN inference partition with load balancing in IoT-edge platform for beyond 5G},
  author={Karjee, Jyotirmoy and Naik, Praveen and Anand, Kartik and Bhargav, Vanamala N},
  journal={Measurement: Sensors},
  volume={23},
  pages={100409},
  year={2022},
  publisher={Elsevier}
}

@inproceedings{hu2022edgeflow,
  title={Distributed inference with deep learning models across heterogeneous edge devices},
  author={Hu, Chenghao and Li, Baochun},
  booktitle={IEEE INFOCOM 2022-IEEE Conference on Computer Communications},
  pages={330--339},
  year={2022},
  organization={IEEE}
}

@article{xiang2021yolocompute,
  title={You only look once, but compute twice: Service function chaining for low-latency object detection in softwarized networks},
  author={Xiang, Zuo and Seeling, Patrick and Fitzek, Frank HP},
  journal={Applied Sciences},
  volume={11},
  number={5},
  pages={2177},
  year={2021},
  publisher={MDPI}
}

@article{bozorgchenani2021multi,
  title={Multi-objective computation sharing in energy and delay constrained mobile edge computing environments},
  author={Bozorgchenani, Arash and Mashhadi, Farshad and Tarchi, Daniele and Monroy, Sergio A Salinas},
  journal={IEEE transactions on mobile computing},
  volume={20},
  number={10},
  pages={2992--3005},
  year={2020},
  publisher={IEEE}
}

@article{fan2025latency,
  title={Latency-aware joint task offloading and energy control for cooperative mobile edge computing},
  author={Fan, Weibei and Xiao, Fu and Pan, Yao and Chen, Xiaobai and Han, Lei and Yu, Shui},
  journal={IEEE Transactions on Services Computing},
  year={2025},
  publisher={IEEE}
}

@article{hmimz2021biobjective,
  title={Bi-objective optimization for multi-task offloading in latency and radio resources constrained mobile edge computing networks},
  author={Hmimz, Youssef and Chanyour, Tarik and El Ghmary, Mohamed and Cherkaoui Malki, Mohammed Oucamah},
  journal={Multimedia Tools and Applications},
  volume={80},
  number={11},
  pages={17129--17166},
  year={2021},
  publisher={Springer}
}

@inproceedings{choi2022gpulet,
  title={Serving heterogeneous machine learning models on $\{$Multi-GPU$\}$ servers with $\{$Spatio-Temporal$\}$ sharing},
  author={Choi, Seungbeom and Lee, Sunho and Kim, Yeonjae and Park, Jongse and Kwon, Youngjin and Huh, Jaehyuk},
  booktitle={2022 USENIX Annual Technical Conference (USENIX ATC 22)},
  pages={199--216},
  year={2022}
}

@inproceedings{ju2022proactive,
  title={Proactive autoscaling for edge computing systems with kubernetes},
  author={Ju, Li and Singh, Prashant and Toor, Salman},
  booktitle={Proceedings of the 14th IEEE/ACM International Conference on Utility and Cloud Computing Companion},
  pages={1--8},
  year={2021}
}

@inproceedings{spatharakis2022autoscaling,
  title={Distributed resource autoscaling in kubernetes edge clusters},
  author={Spatharakis, Dimitrios and Dimolitsas, Ioannis and Vlahakis, Eleftherios and Dechouniotis, Dimitrios and Athanasopoulos, Nikolaos and Papavassiliou, Symeon},
  booktitle={2022 18th International Conference on Network and Service Management (CNSM)},
  pages={163--169},
  year={2022},
  organization={IEEE}
}

\end{document}